\documentclass[seceq,supplement]{ptptex}

\usepackage{graphicx}
\usepackage{wrapft}
\usepackage{epsfig}

\def\mib#1{\hbox{\boldmath $#1$}}
\def\mbf#1{{\mib #1}}
\def\cal#1{{\mathcal #1}}

\def\bp{{\mbf p}}

\def\br{{\mbf r}}

\def\CA{{\cal A}}

\def\CP{{\cal P}}

\def\SM3{\Sigma N (3/2)}
\def\SN1{\Sigma N (1/2)}

\def\TS1{\hbox{}^3S_1}
\def\TD1{\hbox{}^3D_1}

\title{Quark-Model Baryon-Baryon Interaction and its Applications
to Hypernuclei \footnote{Talk presented by Y. Fujiwara at the
18th Nishinomiya Yukawa Memorial Symposium on Strangeness
in Nuclear Matter, 4 - 5 December 2003, Nishinomiya, Japan.
}}

\author{
Yoshikazu \textsc{Fujiwara}, Choki \textsc{Nakamoto}$^{*}$,
Yasuyuki \textsc{Suzuki}$^{**}$,
Michio \textsc{Kohno}$^{***}$
and Kazuya \textsc{Miyagawa}$^{****}$
}

\inst{
Department of Physics, Kyoto University,
Kyoto 606-8502, Japan \\
$\hbox{}^{*}$Suzuka National College of Technology,
Suzuka 510-0294, Japan \\
$\hbox{}^{**}$Department of Physics, Niigata University,
Niigata 950-2181, Japan \\
$\hbox{}^{***}$Physics Division, Kyushu Dental College,
Kitakyushu 803-8580, Japan \\
$\hbox{}^{****}$Department of Applied Physics,
Okayama Science University, Okayama 700-0005, Japan
}

\abst{
The quark-model baryon-baryon interaction fss2, 
proposed by the Kyoto-Niigata group, is a unified model
for the complete baryon
octet ($B_8=N$, $\Lambda$, $\Sigma$ and $\Xi$),
which is formulated in a framework
of the $(3q)$-$(3q)$ resonating-group
method (RGM) using the spin-flavor $SU_6$ quark-model wave functions
and effective meson-exchange potentials at the quark level.
Model parameters are determined to reproduce properties
of the nucleon-nucleon system and the low-energy
cross section data for the hyperon-nucleon scattering.
Due to the several improvements including the introduction
of vector-meson exchange potentials,
fss2 has achieved very accurate description
of the $NN$ and $YN$ interactions,
comparable to various one-boson exchange potentials.
We review the essential features of fss2 and our previous model
FSS, and their predictions to few-body systems
in confrontation with the available experimental data.
Some characteristic features of the $B_8 B_8$ interactions
with the higher strangeness, $S=-2,~-3,~-4$, predicted
by fss2 are discussed.
These quark-model interactions are now applied 
to realistic calculations of few-body systems
in a new three-cluster Faddeev formalism which
uses two-cluster RGM kernels.
As for the few-body systems,
we discuss the three-nucleon bound states,
the $\Lambda NN$-$\Sigma NN$ system for the hypertriton,
the $\alpha \alpha \Lambda$ system
for $\hbox{}^9_{\Lambda}\hbox{Be}$,
and the $\Lambda \Lambda \alpha$ system
for $\hbox{}^{\ 6}_{\Lambda \Lambda} \hbox{He}$.
}

\begin{document}

\maketitle

\section{Introduction}

An important purpose of studying baryon-baryon interactions
in the quark model (QM) is to obtain the most accurate understanding
of the fundamental strong interactions in a natural picture,
in which the short-range part of the interaction is
relevantly described by the quark-gluon degrees of freedom
and the medium- and long-range parts by dominated
meson-exchange processes.
In the spin-flavor $SU_6$ QM, the baryon-baryon interactions
for all the octet baryons ($B_8=N,~\Lambda,~\Sigma$ and $\Xi$) are
consistently treated with the well-known
nucleon-nucleon ($NN$) interaction.
We have recently proposed a comprehensive QM description
of general baryon-octet baryon-octet ($B_8 B_8$) interactions,
which is formulated in the $(3q)$-$(3q)$ resonating-group
method (RGM) using the spin-flavor $SU_6$ QM wave functions,
a colored version of the one-gluon exchange Fermi-Breit interaction,
and effective meson-exchange potentials (EMEP's) acting
between quarks.\cite{FU96a}$^{\hbox{-}}$\cite{FU01b}
The early version, the model FSS,\cite{FU96a}$^{\hbox{-}}$\cite{FJ98}
includes only the scalar (S) and pseudoscalar (PS) meson-exchange
potentials as the EMEP's,
while the renovated one fss2 \cite{FU02a,FU01b} introduces
also the vector (V) meson-exchange potentials
and the momentum-dependent Bryan-Scott terms for the S and V mesons.
Owing to these improvements, the model fss2
in the $NN$ sector has attained
the accuracy comparable to that of one-boson exchange
potentials (OBEP's).

These QM potentials can now be used for various types of
many-body calculations,
which include the $G$-matrix calculations \cite{KO00}
of baryonic matter and the Faddeev calculations of few-baryon
systems.\cite{TR02}
For this purpose, we have recently developed
a new three-cluster formalism
which uses two-cluster RGM kernels explicitly.\cite{TRGM}
The proposed equation eliminates
three-cluster redundant components
by the orthogonality of the total wave function
to the pairwise two-cluster Pauli-forbidden states.
The explicit energy dependence inherent
in the exchange RGM kernel is self-consistently determined.
This equation is entirely equivalent to the Faddeev equation
which uses a modified singularity-free $T$-matrix (which we call
the RGM $T$-matrix) generated from the two-cluster RGM kernel.
We first applied this formalism to a three-dineutron system and
the $3\alpha$ system, and obtained a complete agreement between
the Faddeev calculations and variational calculations which
use the translationally invariant
harmonic-oscillator (h.o.) basis.\cite{TRGM,RED}
Here we apply the formalism to the Faddeev calculations
of the three-nucleon ($3N$) bound state,\cite{TR02} and
the $\Lambda NN$-$\Sigma NN$ system for the hypertriton,
as well as the $\alpha \alpha \Lambda$
and $\Lambda \Lambda \alpha$ systems.

\section{Formulation}

The present model is a low-energy effective model which introduces
some of the essential features of QCD characteristics.
The color degree of freedom of quarks is explicitly
incorporated into the non-relativistic
spin-flavor $SU_6$ quark model,
and the full antisymmetrization of quarks is carried out
in the RGM formalism. The gluon exchange effect is represented
in the form of the quark-quark interaction.
The confinement potential is a phenomenological $r^2$-type potential,
which has a favorable feature that it does not contribute to the
baryon-baryon interactions. We use a color analogue
of the Fermi-Breit (FB) interaction, motivated from the dominant
one-gluon exchange process in conjunction with the asymptotic
freedom of QCD.
We postulate that the short-range part of the baryon-baryon
interaction is well described by the quark degree of freedom.
This includes the short-range repulsion
and the spin-orbit force, both of which are
successfully described by the FB interaction.
On the other hand, the medium range attraction
and the long-range tensor force, especially afforded by the pions,
are extremely non-perturbative from the viewpoint of QCD.
These are therefore most relevantly
described by the effective meson exchange potentials (EMEP).
In the previous model called FSS,\cite{FU96a}$^{\hbox{-}}$\cite{FJ98}
only the scalar and pseudoscalar mesons are introduced.
The model fss2 \cite{FU02a,FU01b} is the most advanced model
which also includes the vector-meson exchange EMEP.
The full QM Hamiltonian $H$ consists of the non-relativistic
kinetic-energy term, the phenomenological confinement
potential $U^{\rm Cf}_{ij}$, the colored
version of the full FB interaction $U^{\rm FB}_{ij}$
with explicit quark-mass dependence,
and the EMEP $U^{\Omega}_{ij}$ generated from 
the scalar ($\Omega$=S), pseudoscalar (PS) and vector (V)
meson exchange potentials acting between quarks:

\begin{figure}[ht]
\centerline{\epsfxsize=\textwidth\epsfbox{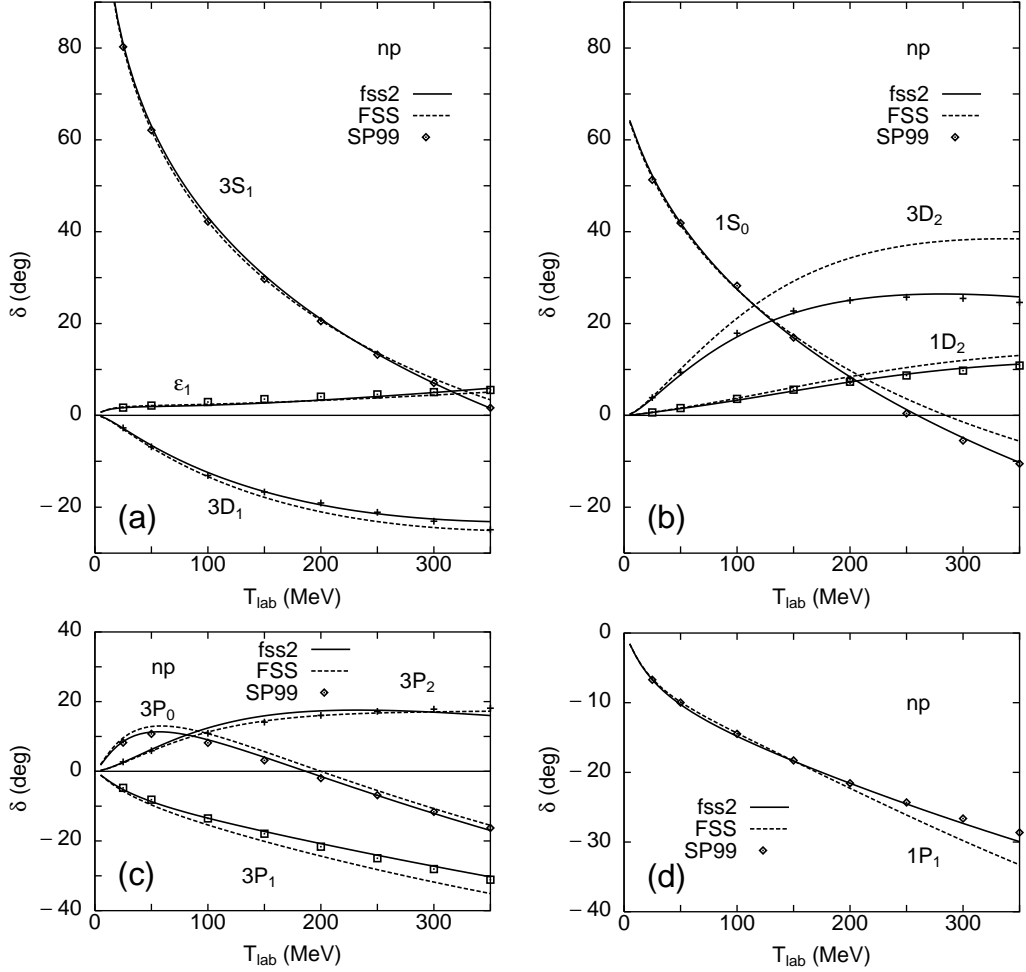}}
\caption{Comparison of $np$ phase-shift parameters
for $J \leq 2$ with the phase-shift analysis SP99
by Arndt {\em et al.} \protect\cite{SAID}
The solid curves are the results by fss2 and
the dotted curves by FSS.}
\label{fig1}
\end{figure}

\begin{eqnarray}
H=\sum^6_{i=1} \left( m_ic^2+{\bp^2_i \over 2m_i}-T_G \right)
+\sum^6_{i<j} \left( U^{\rm Cf}_{ij}+U^{\rm FB}_{ij}
+U^{\rm S}_{ij}
+U^{\rm PS}_{ij}
+U^{\rm V}_{ij} \right).\ 
\label{fm1}
\end{eqnarray}
The RGM equation for the relative-motion
wave function $\chi(\br)$ reads
\begin{eqnarray}
\langle \phi(3q)\phi(3q)\vert E-H \vert
\CA\left\{\phi(3q)\phi(3q)\chi(\br)\right\}\rangle=0.
\label{fm2}
\end{eqnarray}
We solve this RGM equation in the momentum
representation.\cite{LSRGM}
If we rewrite the RGM equation in the form
of the Schr{\"o}dinger-type
equation as $\left[\varepsilon-H_0-V_{\rm RGM}(\varepsilon)\right]$
$\times \chi(\br)=0$, the potential term,
$V_{\rm RGM}(\varepsilon)=V_{\rm D}+G+\varepsilon K$,
becomes nonlocal and energy dependent.
Here $V_{\rm D}$ represents the direct potential of EMEP's,
$G$ includes all the exchange kernels for the interaction
and kinetic-energy terms, and $K$ is the exchange
normalization kernel. 
We calculate the plane-wave matrix elements
of $V_{\rm RGM}(\varepsilon)$, and set up with
the Lippmann-Schwinger equation of the RGM $T$-matrix.
This approach is convenient to proceed to
the $G$-matrix calculations. \cite{KO00,GRGM}
Faddeev calculations using these exchange kernels are also possible
with some special considerations of the Pauli forbidden
states. \cite{TRGM,RED}

\begin{figure}[t] 
\centerline{\epsfxsize=\textwidth\epsfbox{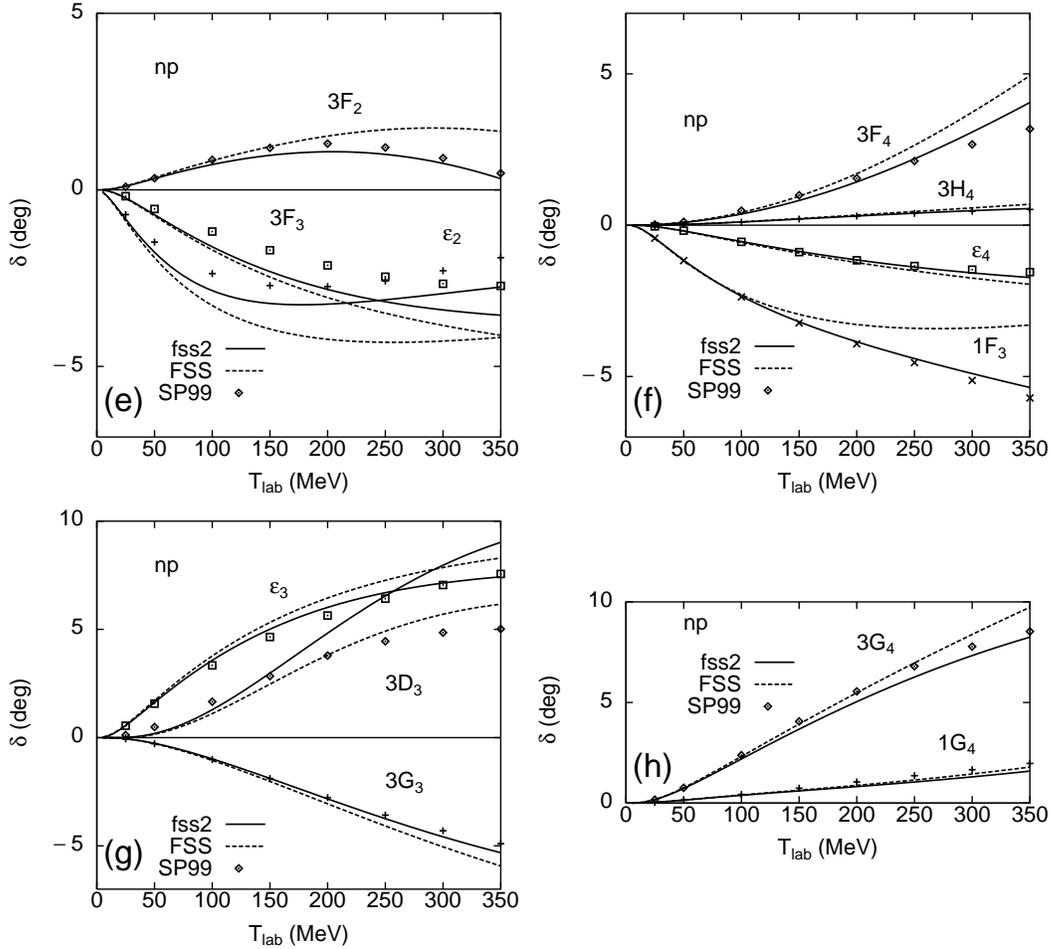}}
\caption{Same as Fig. 1 but for the higher partial waves
with $J \leq 4$.}
\label{fig2}
\end{figure}

\section{\mib{B_8 B_8} interactions by fss2 and FSS}

\subsection{$NN$ properties and $G$-matrix calculations
of nuclear matter}

\begin{figure}[t] 
\vspace{2mm}
\begin{minipage}[h]{0.48\textwidth}
\centerline{\epsfxsize=\textwidth\epsfbox{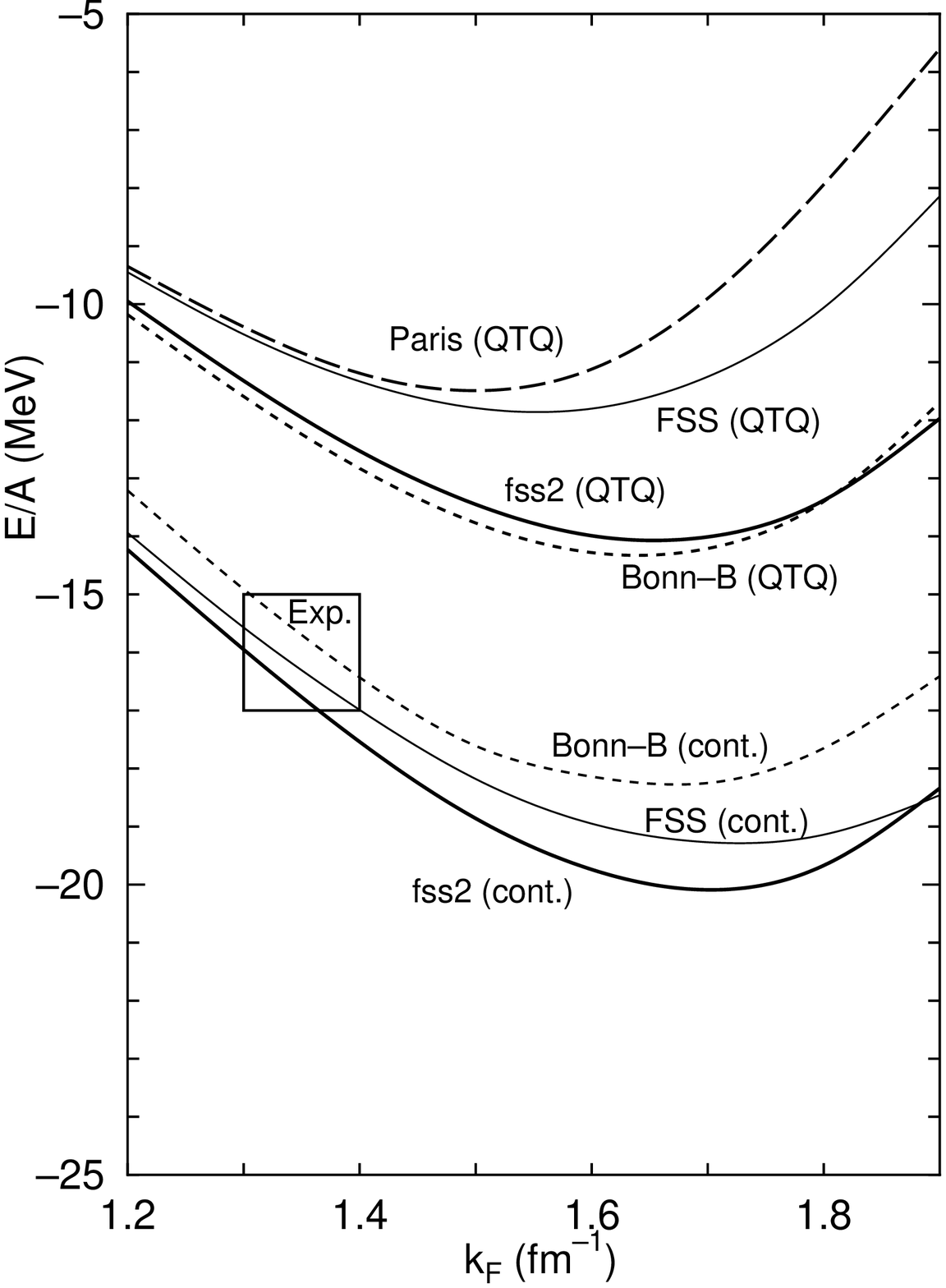}} 
\vspace{-5mm}
\caption{
Nuclear matter saturation curves obtained by fss2 and FSS,
together with the results of the Paris
potential\protect\cite{PARI} and
the Bonn model-B potential\protect\cite{MA89}.
The choice of the intermediate spectra
is specified by ``QTQ" and ``cont."
The result for the Bonn-B potential in the continuous
choice is taken from the non-relativistic calculation
in Ref.~\protect\citen{BM90}.}
\label{fig3}
\end{minipage}
\hfill
\begin{minipage}[h]{0.48\textwidth}
\centerline{\epsfxsize=\textwidth\epsfbox{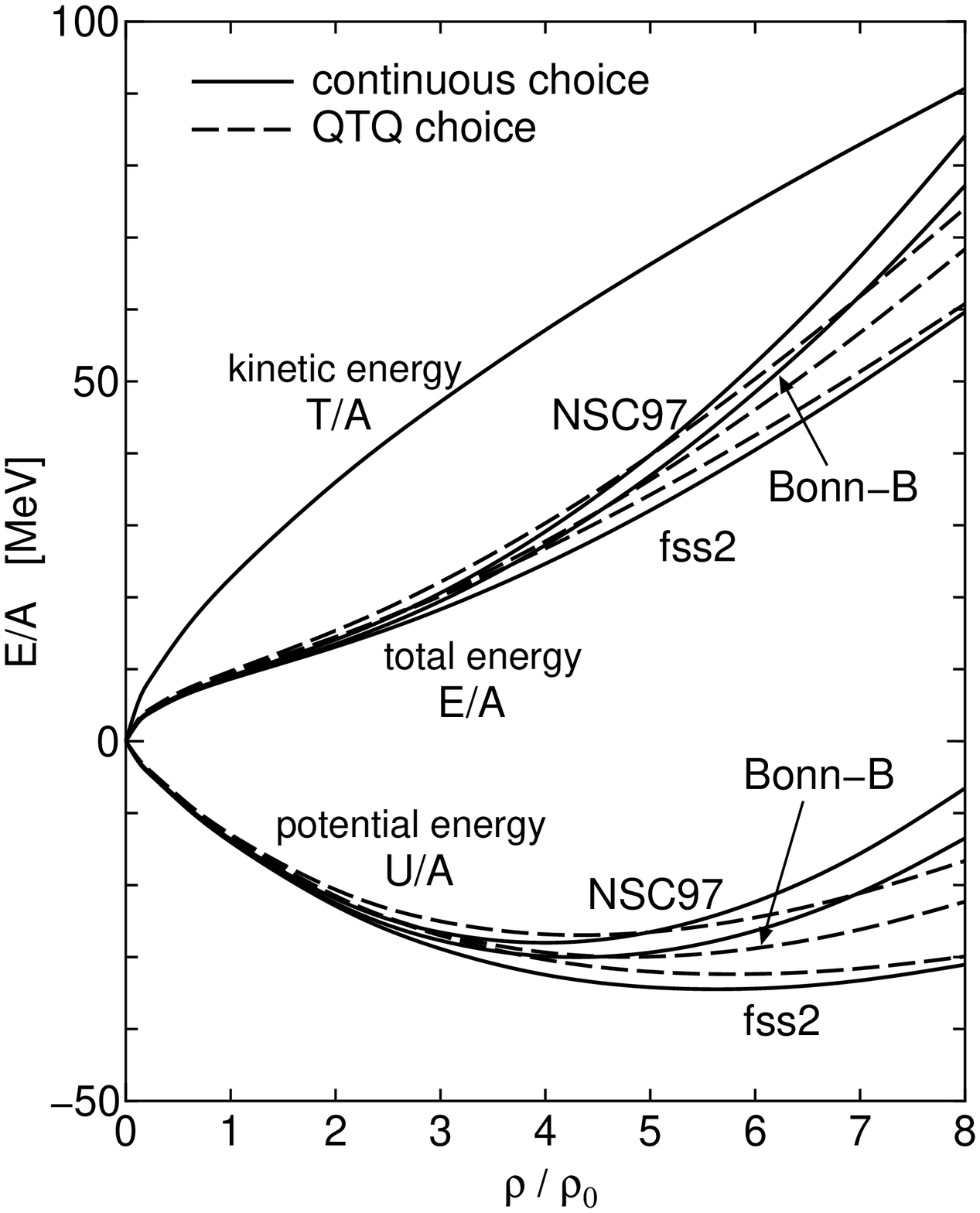}} 
\caption{Same as Fig.~\protect\ref{fig3} but for the neutron matter
saturation curve predicted by fss2.
The normal density $\rho_0$ corresponds
to the Fermi momentum $k^n_F=1.35~\hbox{fm}^{-1}$.
The results of the Nijmegen soft-core
potential (NSC97) \protect\cite{NSC97} and the
Bonn-B potential \protect\cite{MA89} are also shown for comparison.}
\label{fig4}
\end{minipage}
\end{figure}

\begin{table}[b]
\caption{The deuteron properties
by fss2 and FSS in the isospin basis.
The results by the Bonn-B potential \protect\cite{MA89} are
also shown for comparison. A small difference in FSS from Table IV
of Ref.~\protect\citen{FU96b} is due to the numerical inaccuracy
in the previous calculation.
The effect of the meson exchange current is not included
in the calculations of $Q_d$ and $\mu_d$.}
\renewcommand{\arraystretch}{1.2}
\setlength{\tabcolsep}{5mm}
\begin{center}
\begin{tabular}{llllll}
\hline
 & FSS & fss2 & Bonn B & Expt. & Ref. \\
\hline
$\epsilon_d$ (MeV) & 2.2561 & 2.2247 & 2.2246
 & $2.224644 \pm 0.000046$
 & \protect\citen{DU83} \\
$P_D$ ($\%$)       & 5.86   & 5.49   & 4.99   & $-$ & \\
$\eta=A_D/A_S$     & 0.0267 & 0.0253 & 0.0264 & $0.0256 \pm 0.0004$
 & \protect\citen{RO90} \\
rms (fm) & 1.963   & 1.960  & 1.968  & $1.9635 \pm 0.0046$
 & \protect\citen{DU83} \\
$Q_d$ (fm$\hbox{}^2$) & 0.283 & 0.270 & 0.278 & $0.2860 \pm 0.0015$
 & \protect\citen{BI79} \\
$\mu_d$ ($\mu_N$) & 0.8464  & 0.8485 & 0.8514 &
$0.857406 \pm 0.000001$ & \protect\citen{Lin65} \\
\hline
\end{tabular}
\end{center}
\label{table1}
\end{table}

Figures 1 and 2 display some important low-partial
wave $NN$ phase-shift
parameters predicted by the model fss2, in comparison
with the phase shift analysis SP99. \cite{SAID}
The previous results by FSS are also shown with the dotted curves.
Due to the inclusion of V mesons, the $NN$ phase shifts
of the fss2 at the non-relativistic energies
up to  $T_{\rm lab}=350~\hbox{MeV}$ are greatly improved,
and now have attained the accuracy almost comparable
to that of OBEP's.  
The good reproduction of the $NN$ phase-shift
parameters in fss2 continues up
to $T_{\rm lab} \sim 600~\hbox{MeV}$,\cite{FU02a}
where the inelasticity of the $S$-matrix becomes appreciable.

For the correct evaluation of the triton binding energy,
it is well known that the proper reproduction
of the deuteron $D$-state probability
and the $\hbox{}^1S_0$ effective range parameters is essential.
Table \ref{table1} shows the deuteron properties predicted
by FSS and fss2, in comparison with the experiment.
The predictions by the Bonn-B potential,
which has a smaller deuteron $D$-state probability, is also
shown for comparison.

\begin{figure}[t] 
\begin{minipage}[h]{0.48\textwidth}
\centerline{\epsfxsize=\textwidth\epsfbox{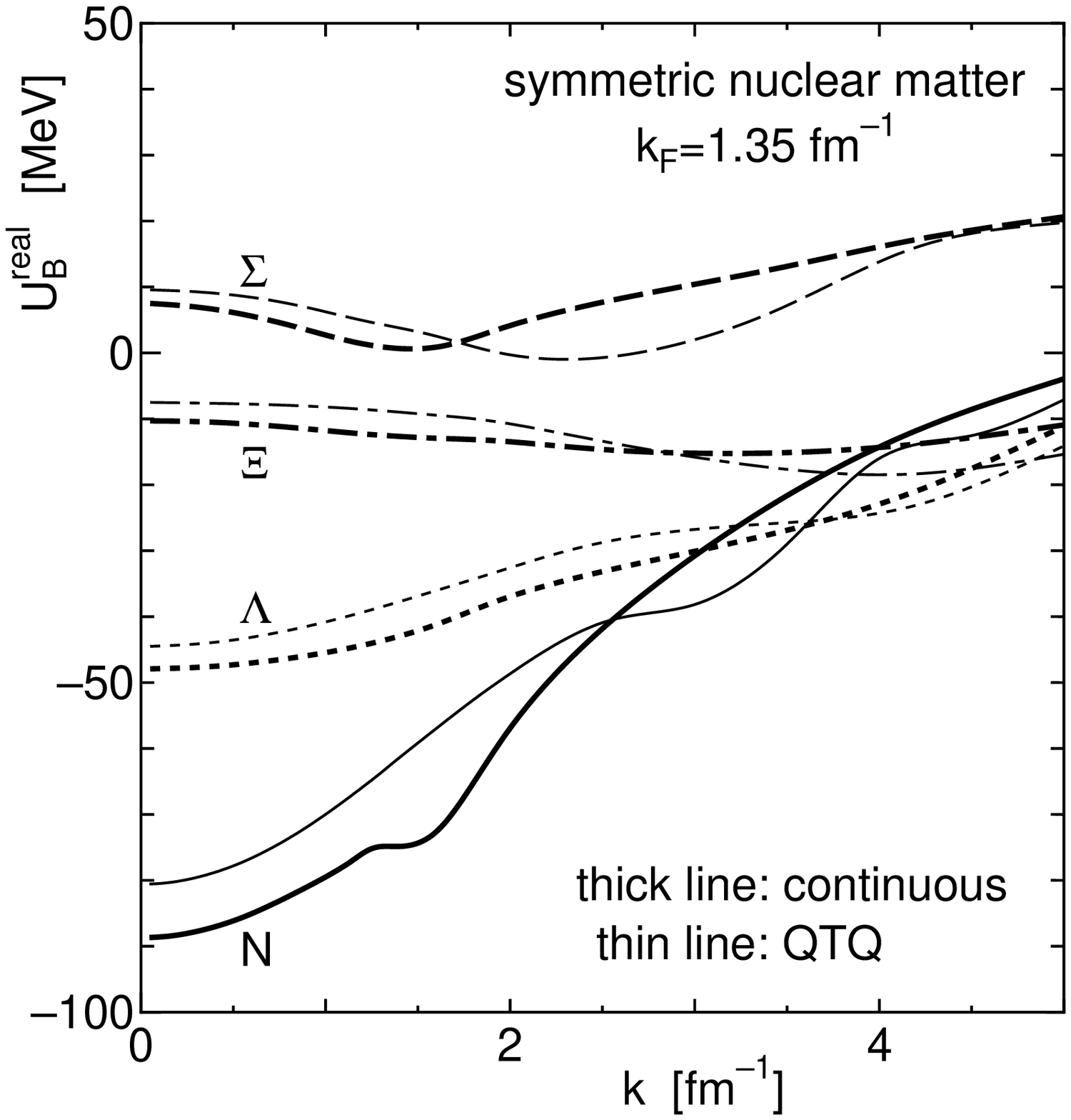}}
\caption{
The s.p. potentials $U_B (k)$ in symmetric
nuclear matter, predicted by fss2, for the normal
density $\rho_0$ with $k_F =1.35~\hbox{fm}^{-1}$.}
\label{fig5-1}
\end{minipage}
\hfill
\begin{minipage}[h]{0.48\textwidth}
\centerline{\epsfxsize=\textwidth\epsfbox{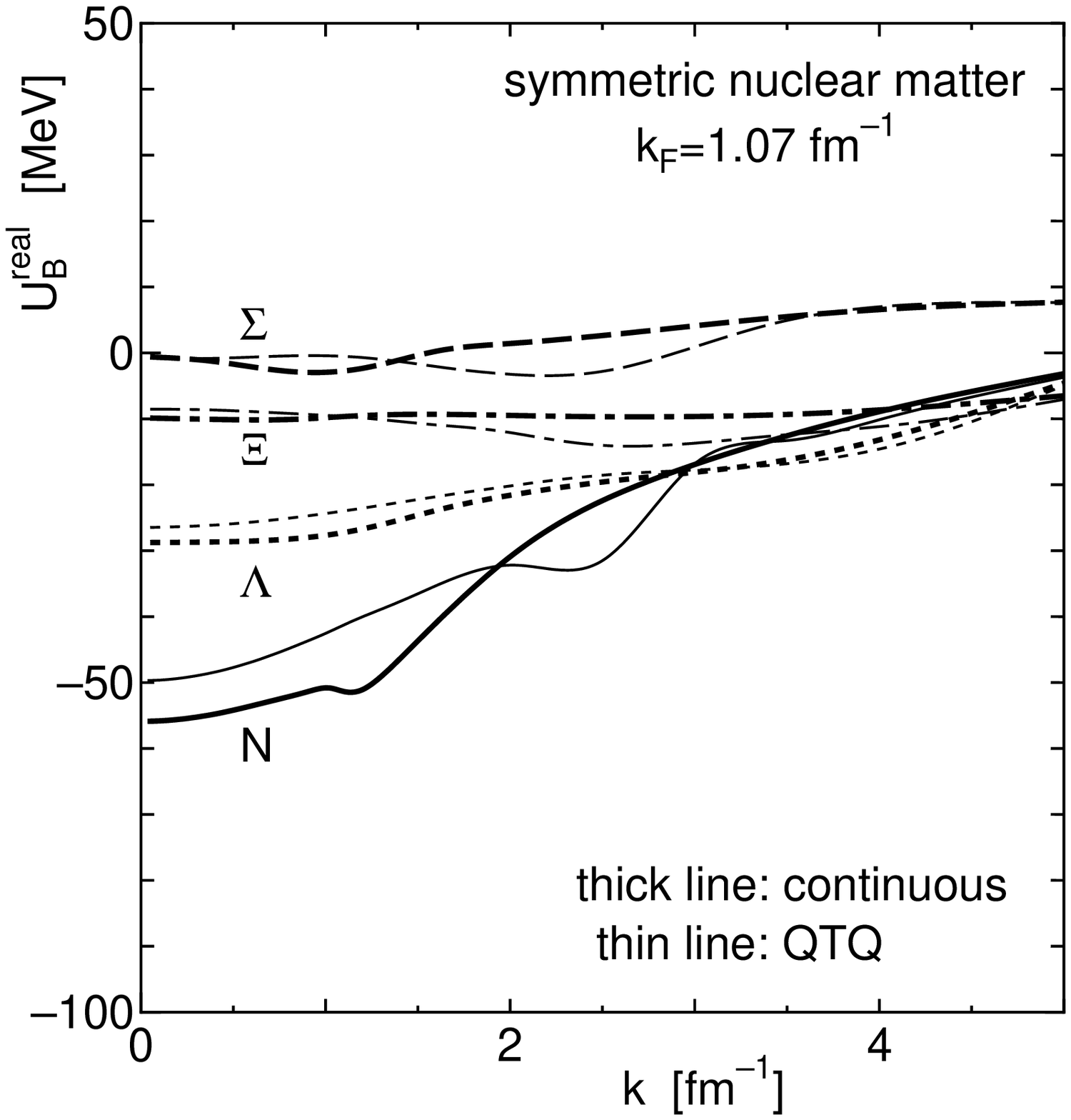}}
\caption{
Same as Fig.~\protect\ref{fig5-1} but for half of the normal
density ($\rho= 0.5\,\rho_0$) with $k_F =1.07~\hbox{fm}^{-1}$.}
\label{fig5-2}
\end{minipage}
\end{figure}

Figure~\ref{fig3} shows saturation curves
of symmetric nuclear matter,
obtained by fss2 and FSS. \cite{KO00}
They depend on the prescription how
one deals with the energy spectrum of the intermediate
single-particle (s.p.) states.
In both cases of the QTQ and the continuous choices,
they are very similar to the predictions by meson exchange
potentials. In the continuous choice, the saturation curves
of fss2 and FSS are very similar
to that of the Bonn-B potential. \cite{MA89}
Unfortunately, our results share the common unsatisfactory 
feature of any non-relativistic models,
that the saturation point does not deviate from the Caester line.

The saturation curve for neutron matter, predicted by fss2,
is also shown in Fig.~\ref{fig4}. The results by the
Nijmegen soft-core potential (NSC97) \cite{NSC97} and the
Bonn-B potential \cite{MA89} are also shown for comparison.
Since the strongly attractive $\hbox{}^3S_1+\hbox{}^3D_1$ channel
does not exist in this case, the total energy per neutron in neutron
matter becomes repulsive in any models. We find that our quark model
potential gives very similar results to the standard meson-exchange
potentials, as long as the $NN$ interaction is concerned.

The s.p. potentials of the nucleon and hyperons
in symmetric nuclear matter are shown in Figs. \ref{fig5-1} 
($k_F=1.35~\hbox{fm}^{-1}$) and \ref{fig5-2}
($k_F=1.07~\hbox{fm}^{-1}$) for the model fss2. 
For the standard Fermi momentum $k_F=1.35~\hbox{fm}^{-1}$,
which corresponds to the normal density $\rho_0$,
these are fairly deep potentials. 
For the comparison with the depth of the s.p. potentials
in the finite nuclei, one has to take a smaller value
for $k_F$ because of the surface effects. 
If we assume $k_F=1.07~\hbox{fm}^{-1}$, the s.p. potentials become
much shallower, and the depth becomes
almost comparable with the empirical values, $-50$ MeV for $N$,
$-30$ MeV for $\Lambda$, and $-10 \sim -14$ MeV for $\Xi$.
The sign of the $\Sigma$ s.p. potential is still controversial.


\begin{figure}[t]
\begin{minipage}{0.49\textwidth}
\centerline{\epsfxsize=\textwidth\epsfbox{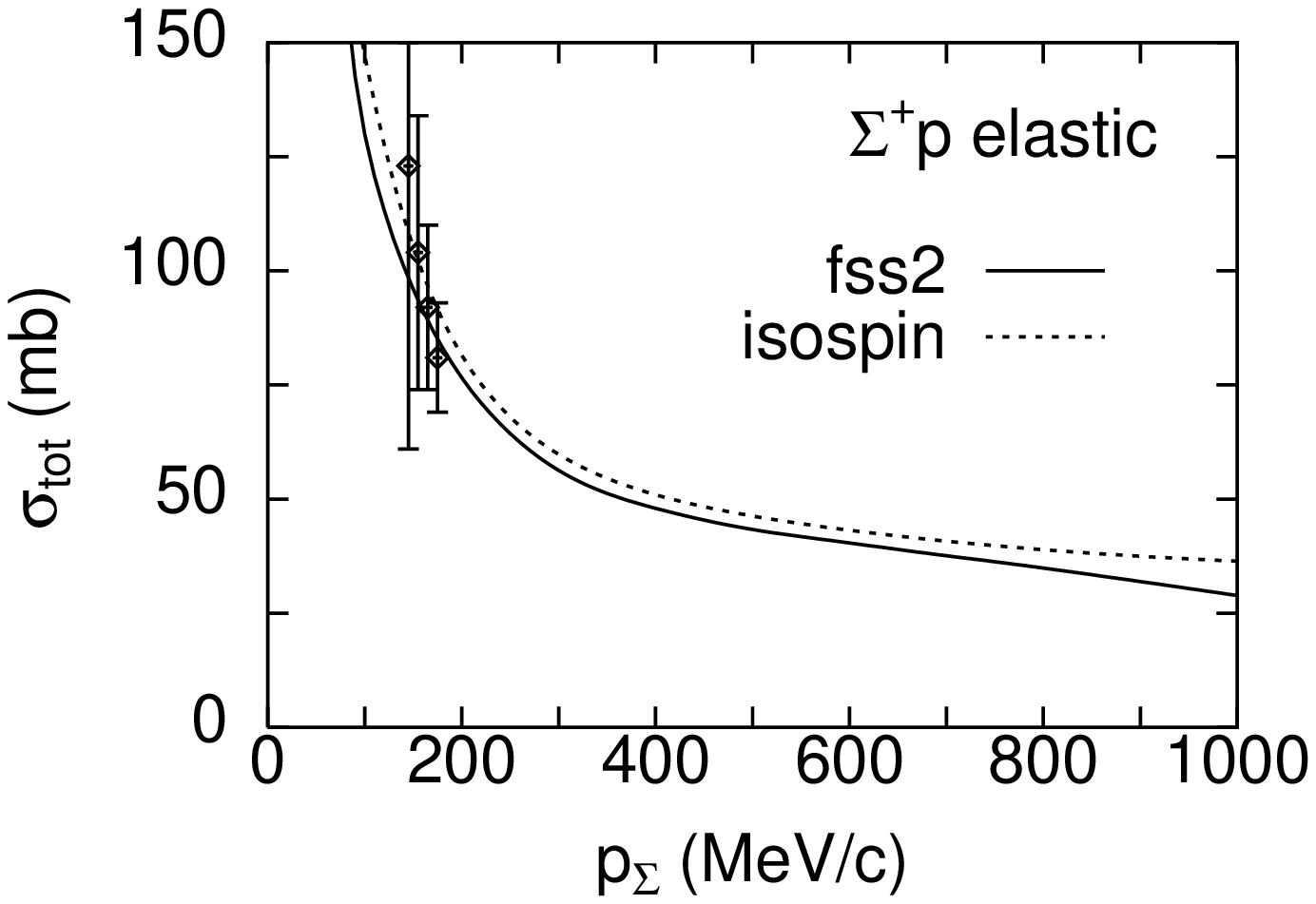}}
\centerline{\epsfxsize=\textwidth\epsfbox{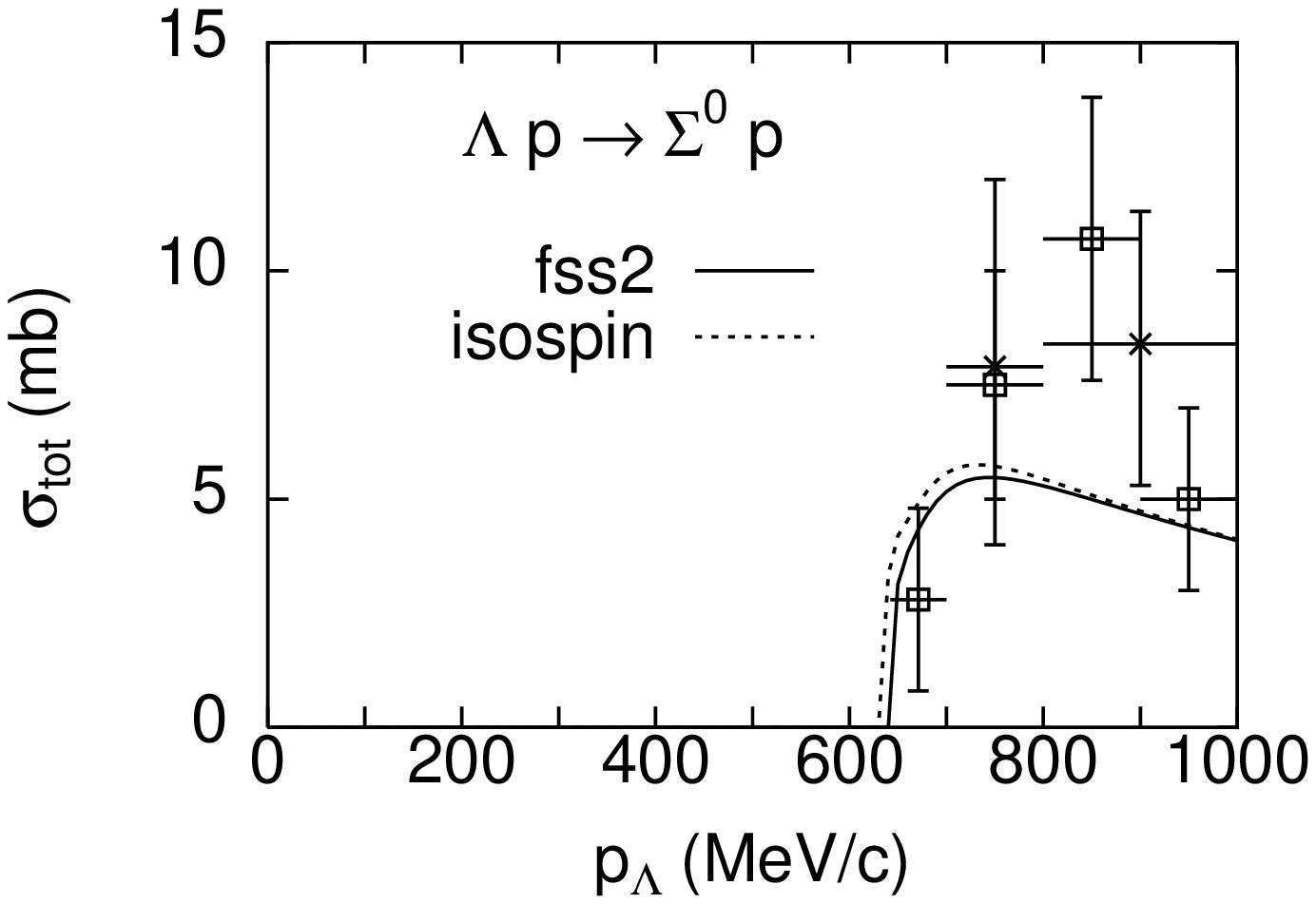}}
\centerline{\epsfxsize=\textwidth\epsfbox{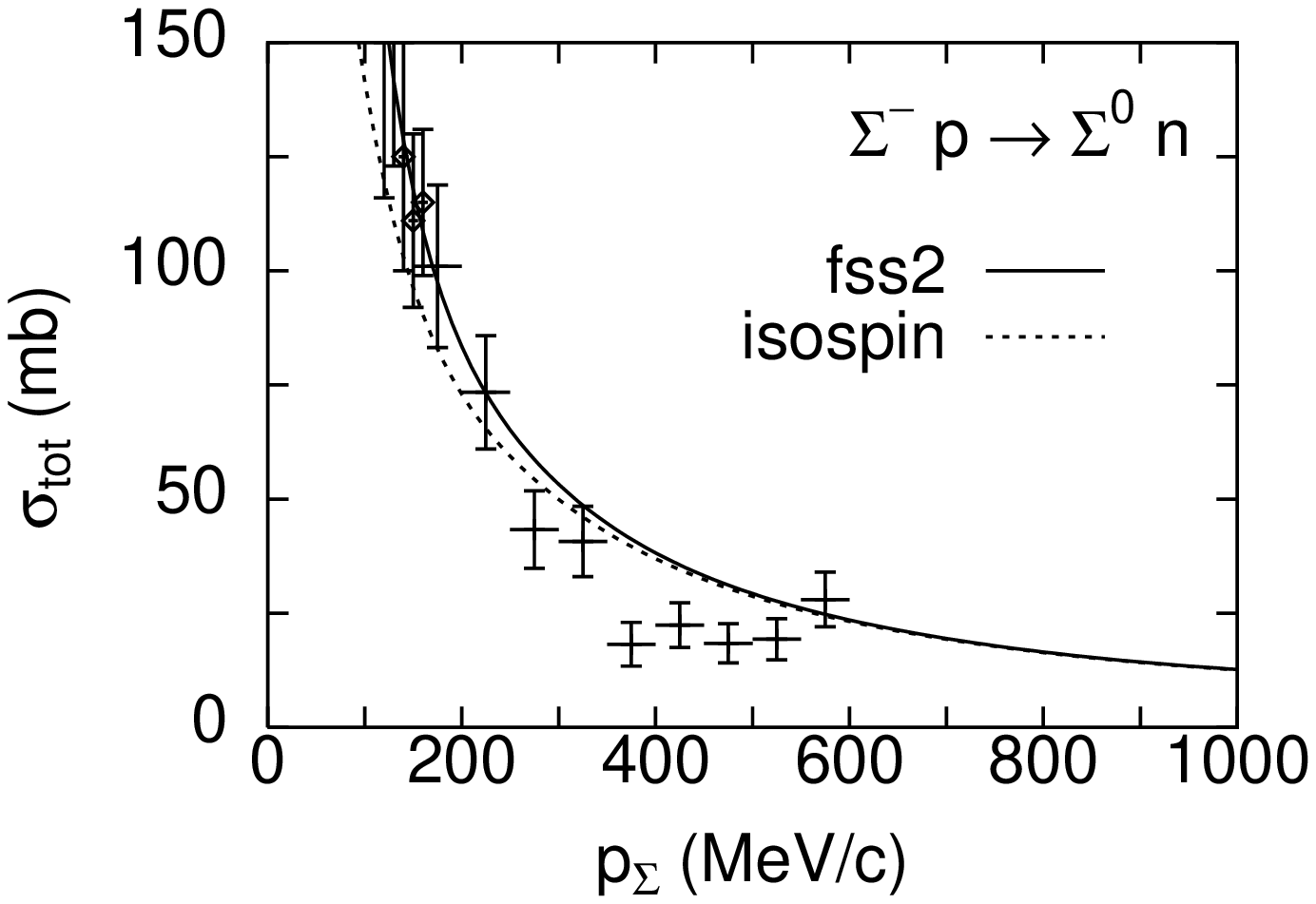}}
\end{minipage}~%
\begin{minipage}{0.49\textwidth}
\centerline{\epsfxsize=\textwidth\epsfbox{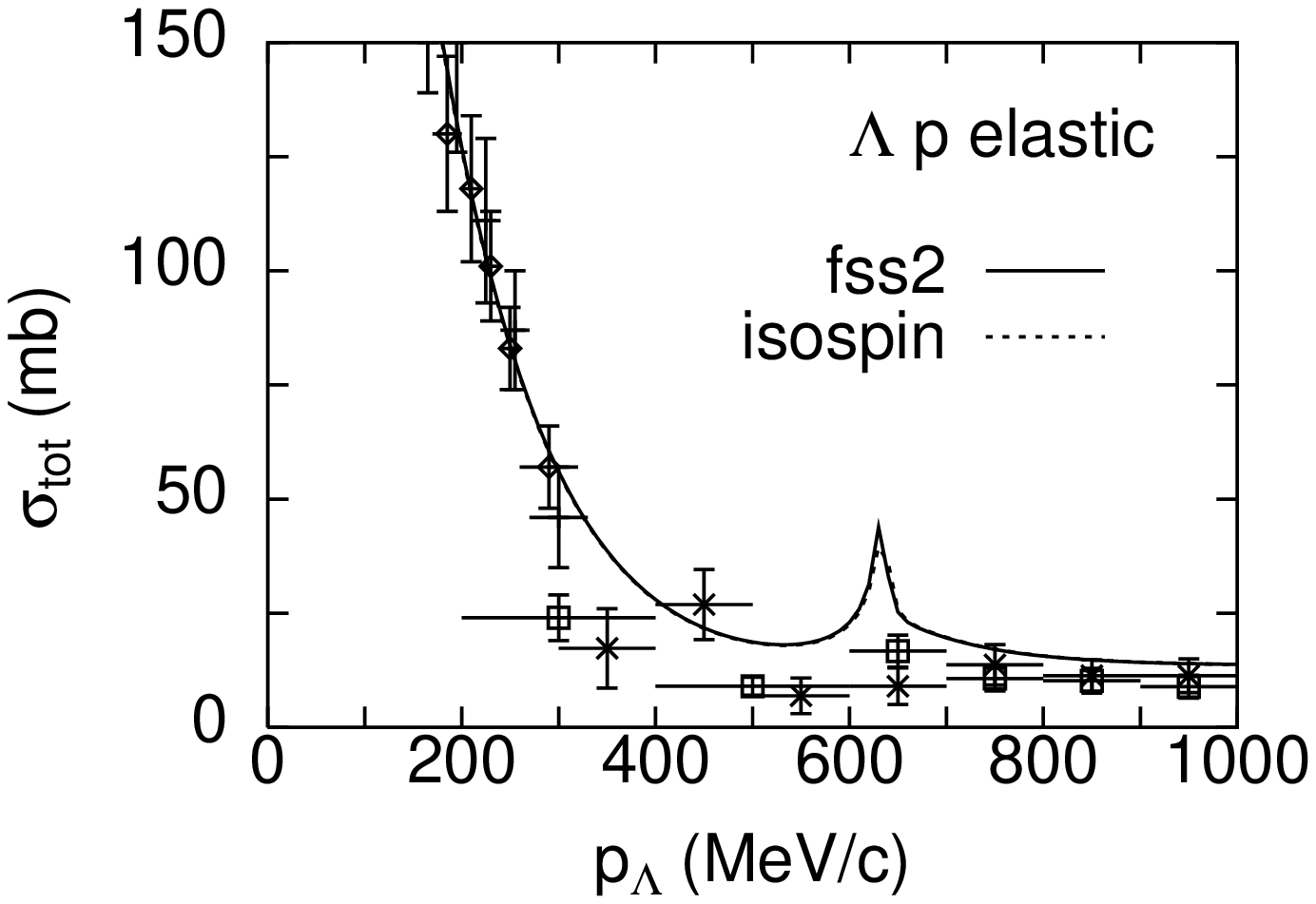}}
\centerline{\epsfxsize=\textwidth\epsfbox{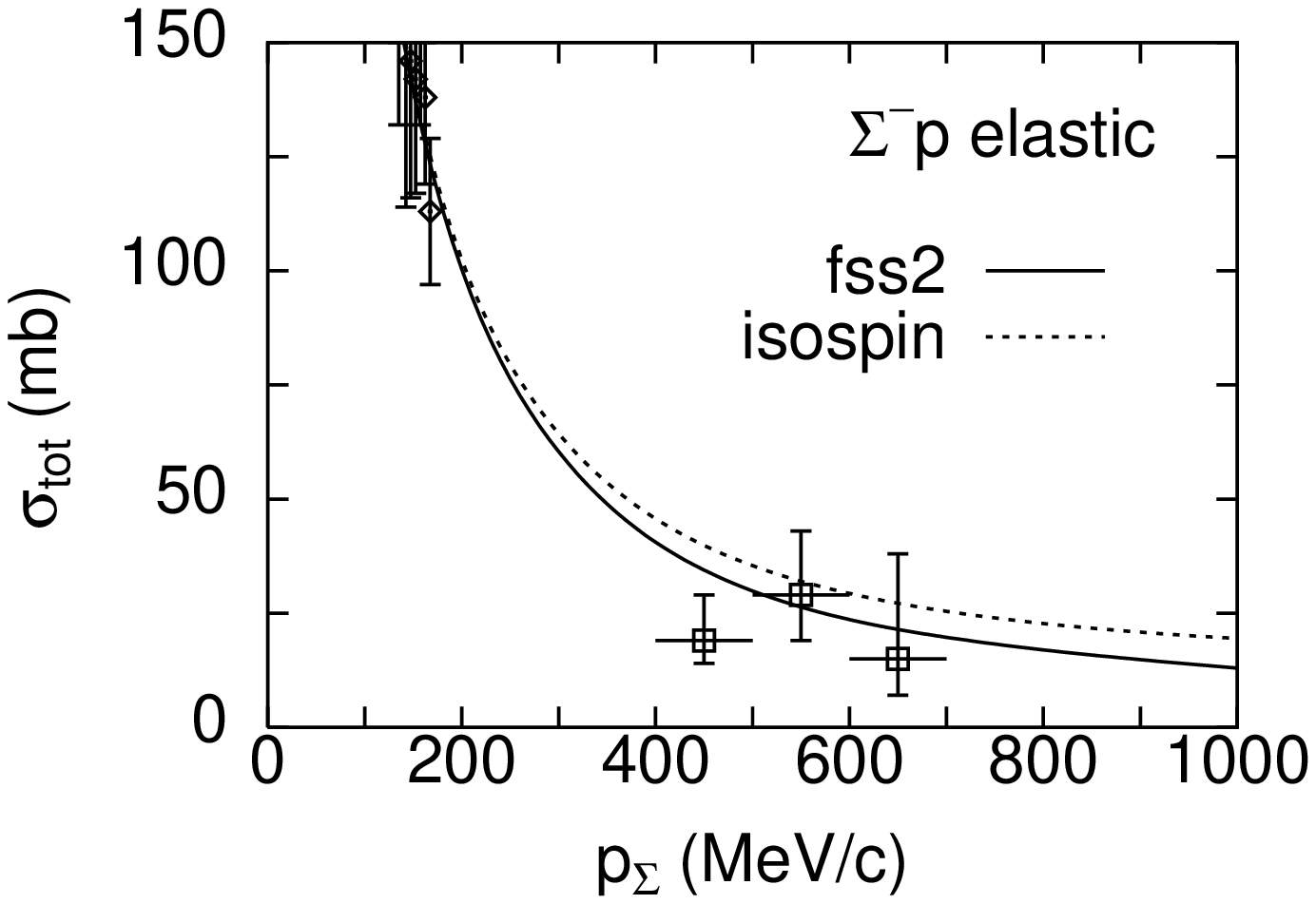}}
\centerline{\epsfxsize=\textwidth\epsfbox{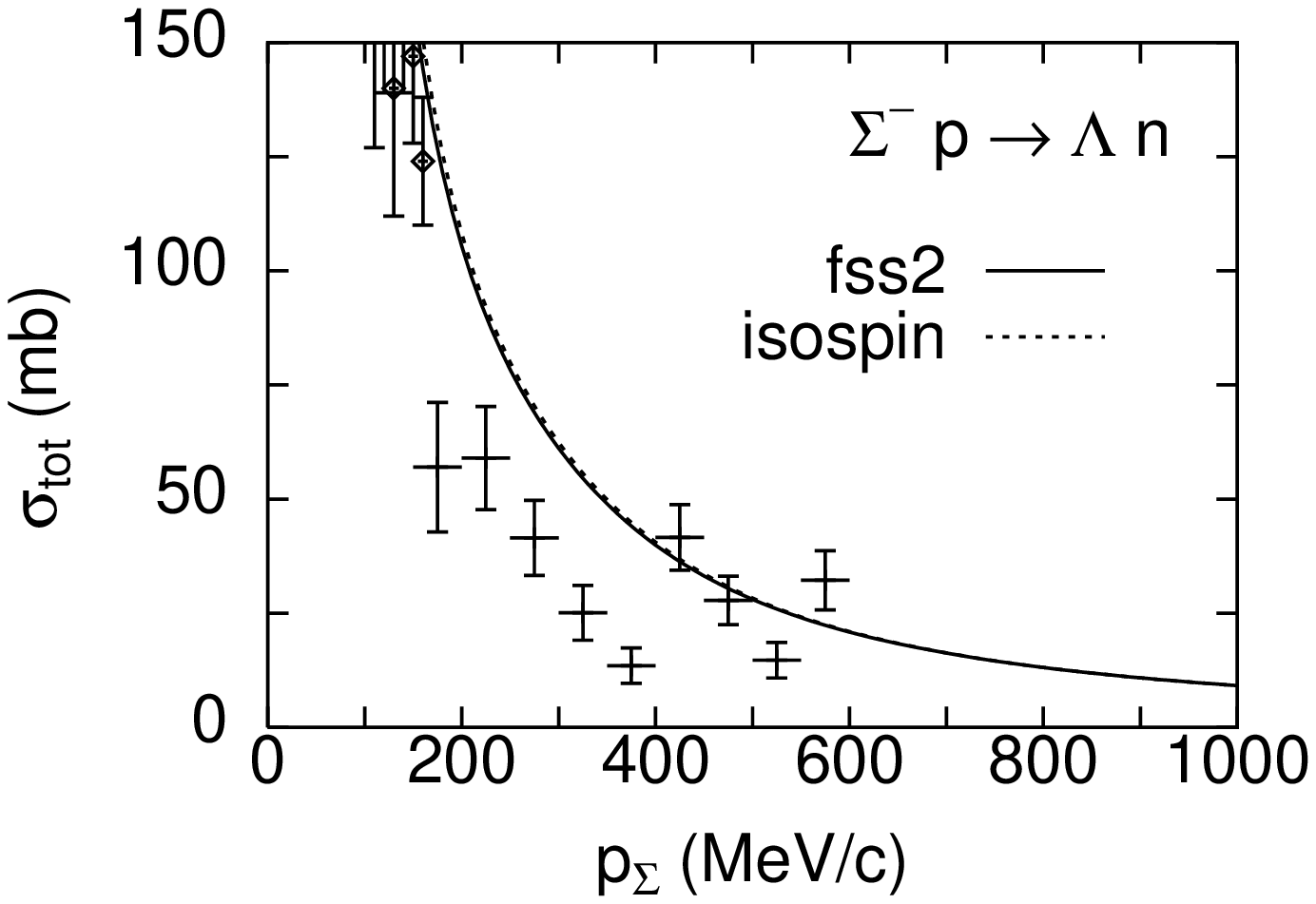}}
\end{minipage}
\caption{Calculated $YN$ total cross sections
compared with the available experimental data.}
\label{fig7}
\end{figure}

\subsection{$YN$ and $YY$ interactions by fss2 and FSS}

The total cross sections of the hyperon-nucleon ($YN$) scattering
predicted by fss2 are compared with the available
experimental data in Fig.~\ref{fig7}.
The ``total'' cross sections for the scattering of charged
particles (i.e., $\Sigma^+ p$ and $\Sigma^- p$ systems) are
calculated by integrating the differential cross sections
over $\cos \theta_{\rm min}=0.5 \sim \cos \theta_{\rm max}=-0.5$.
The solid curves indicate the result in the particle basis,
while the dashed curves in the isospin basis. In the latter case,
the effects of the charge symmetry breaking,
such as the Coulomb effect and the small difference
of the threshold energies for $\Sigma^- p$
and $\Sigma^0 n$ channels, are neglected.
New experimental data for $\Sigma^- p$ elastic total cross sections
at the intermediate energies,
$p_\Sigma=400 \hbox{-} 700~\hbox{MeV}/c$,
measured at KEK,\cite{KON00} are consistent
with the fss2 predictions.

\begin{figure}[t]
\begin{minipage}{0.49\textwidth}
\centerline{\epsfxsize=\textwidth\epsfbox{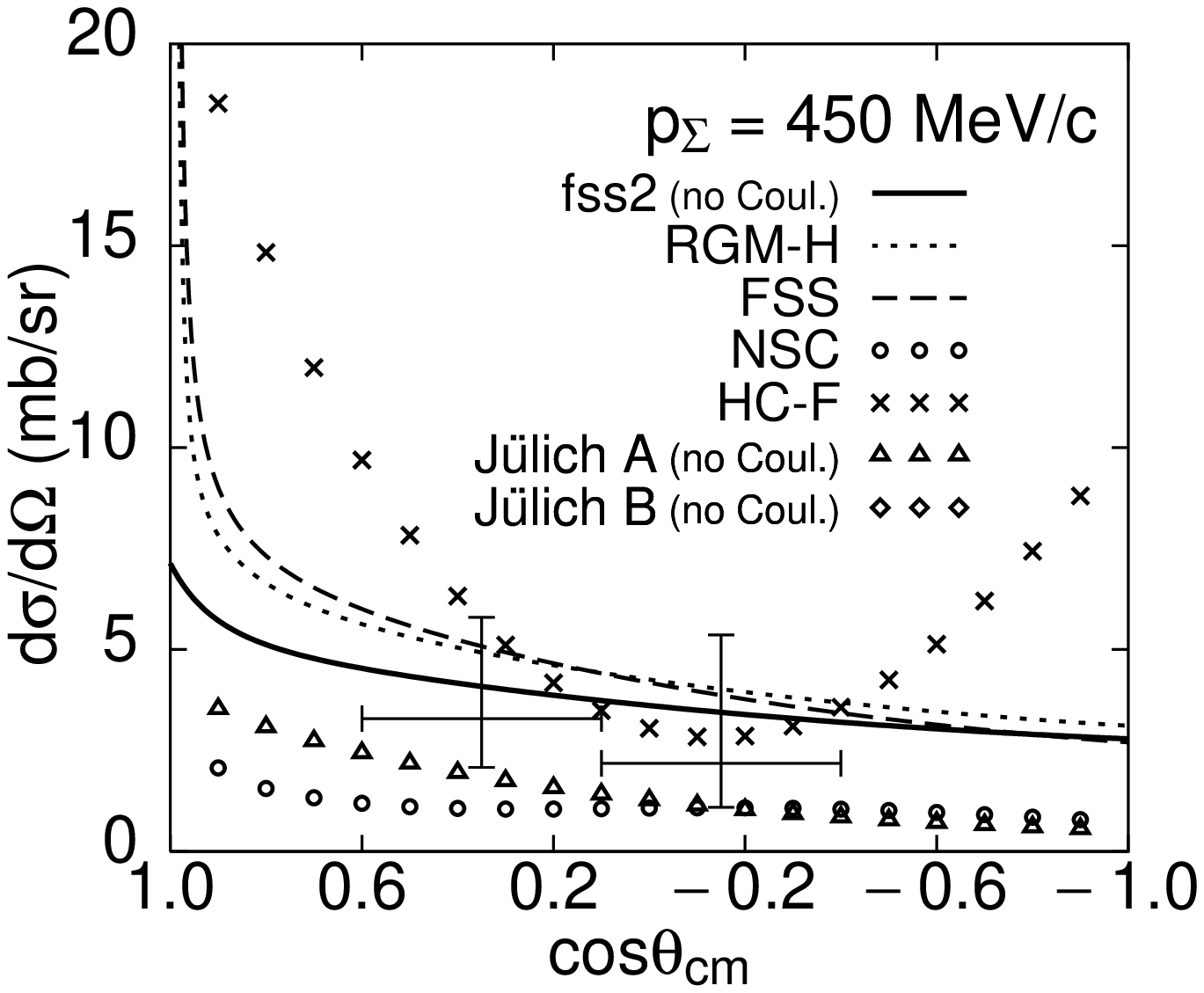}}
\end{minipage}~%
\begin{minipage}{0.48\textwidth}
\vspace{4mm}
\includegraphics[angle=-90,width=\textwidth]{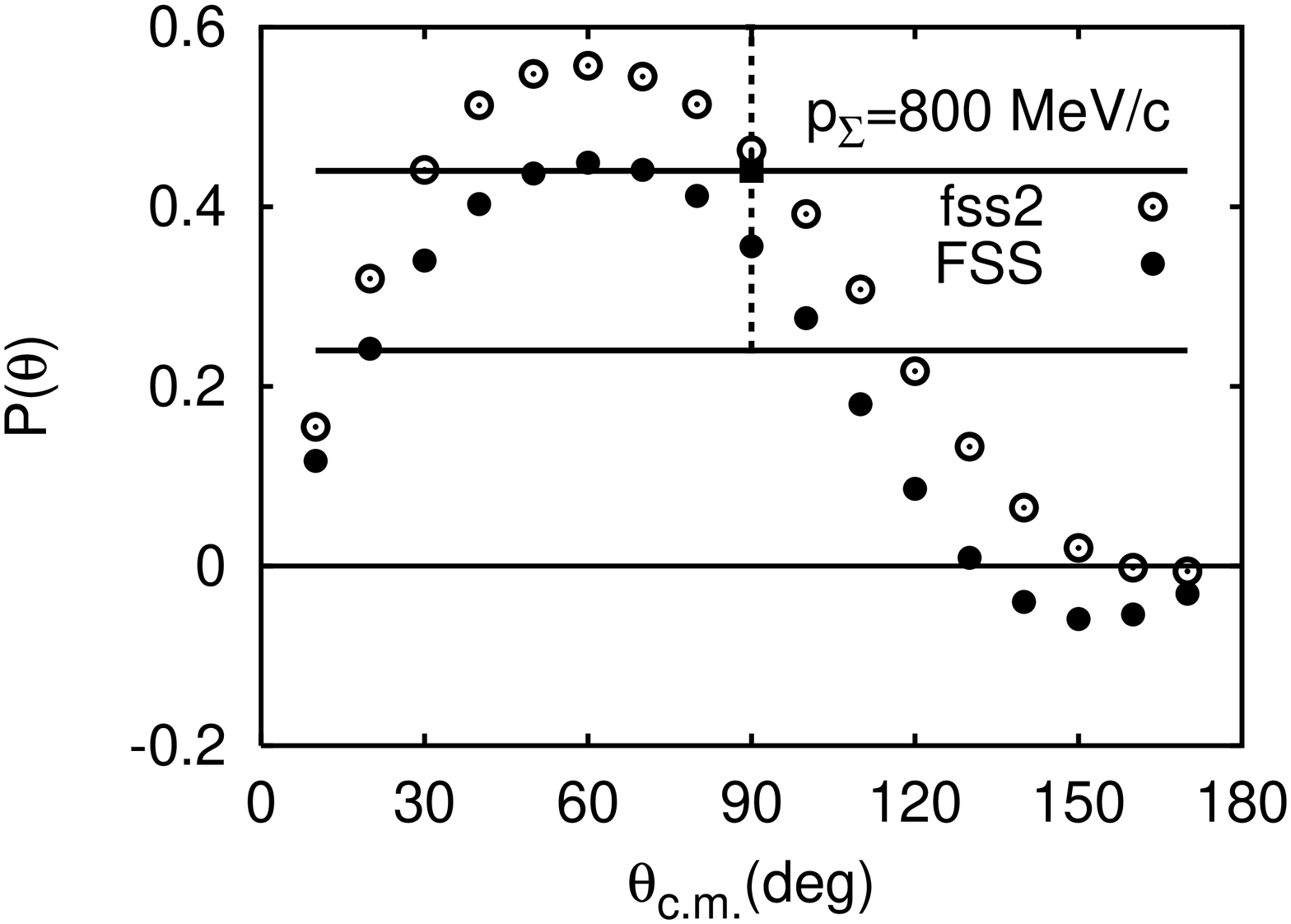}
\end{minipage}
\caption{Left: $\Sigma^+p$ differential cross sections
at $p_\Sigma=450~\hbox{MeV}/c$ predicted by various models.
The experimental data are cited from Ref.~\protect\citen{GO97}.
Right: $\Sigma^+p$ poralization at $p_\Sigma=800~\hbox{MeV}/c$,
predicted by fss2 and FSS. 
The experimental data are cited from Ref.~\protect\citen{KA02}.}
\label{fig8}
\end{figure}

The $\Sigma^+p$ differential cross sections
at the intermediate energy $p_\Sigma=450~\hbox{MeV}/c$ are
compared with the KEK experiment \cite{GO97} in the
left panel of Fig.~\ref{fig8}.
We need more experimental data to increase the statistics,
in order to see which model is the most appropriate.
In the right panel, the polarization observables for
the $\Sigma^+ p$ elastic scattering at $p_\Sigma=800~\hbox{MeV}/c$
are shown for the models fss2 and FSS.
The recent experimental data from KEK-PS 457 \cite{KA02} imply
the asymmetry parameter $a^{\rm exp}=0.44 \pm 0.2$ at $p_\Sigma=800
\pm 200~\hbox{MeV}/c$, which is not inconsistent with our
quark-model predictions although the specific scattering angle
is not possible to be identified.

\begin{figure}[t]
\begin{minipage}{0.49\textwidth}
\includegraphics[angle=-90,width=\textwidth]{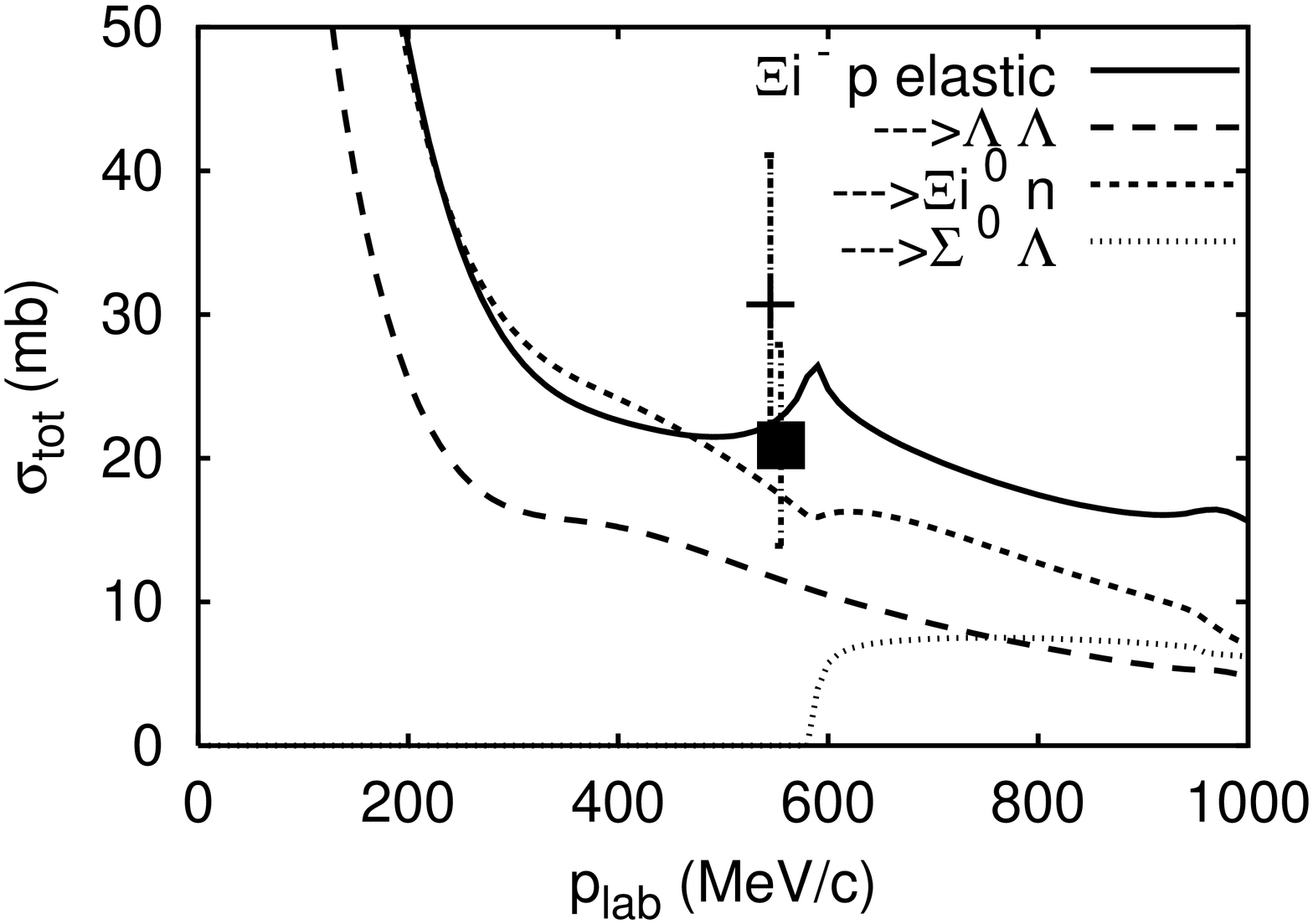}
\includegraphics[angle=-90,width=\textwidth]{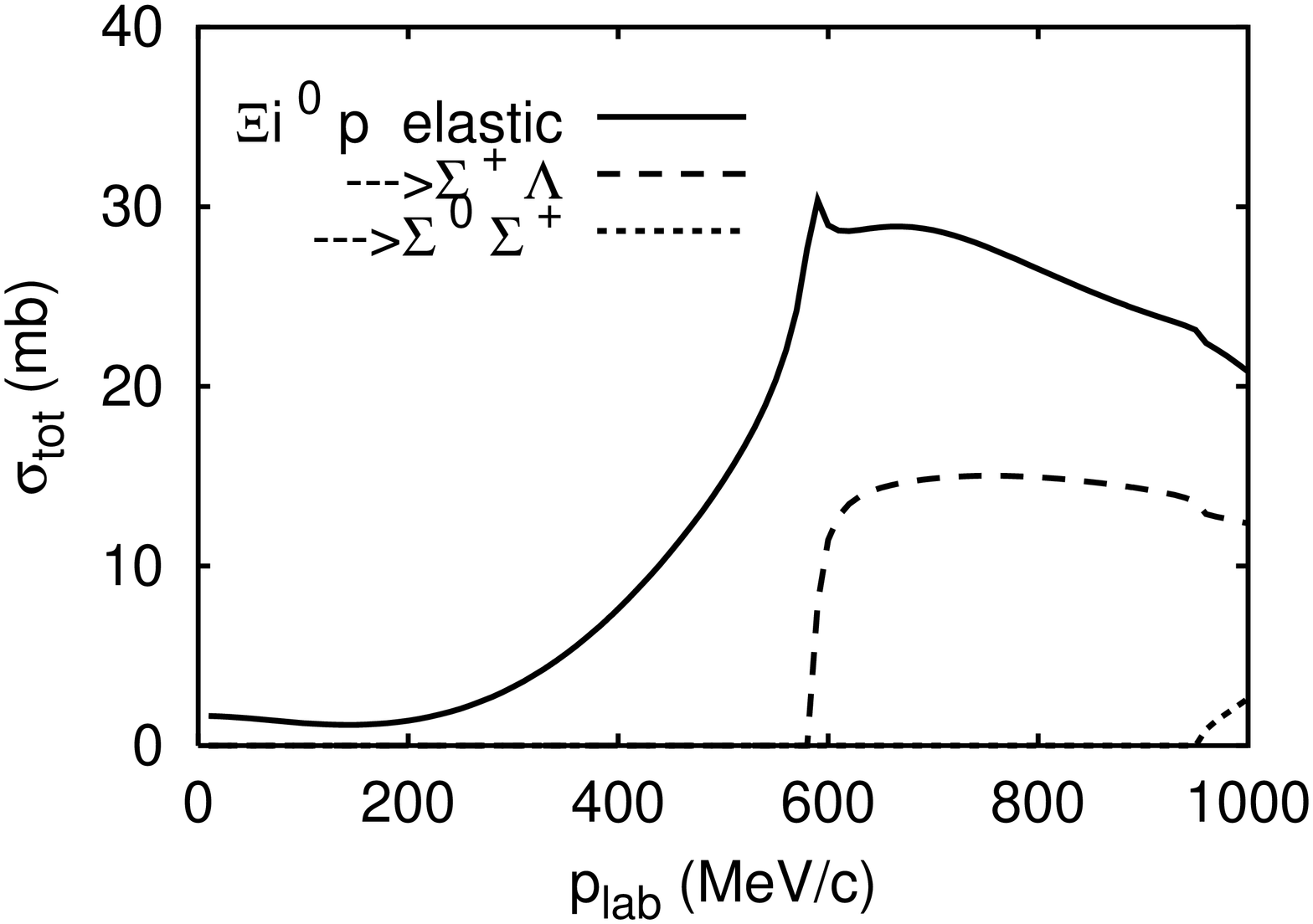}
\end{minipage}~%
\begin{minipage}{0.49\textwidth}
\includegraphics[angle=-90,width=\textwidth]{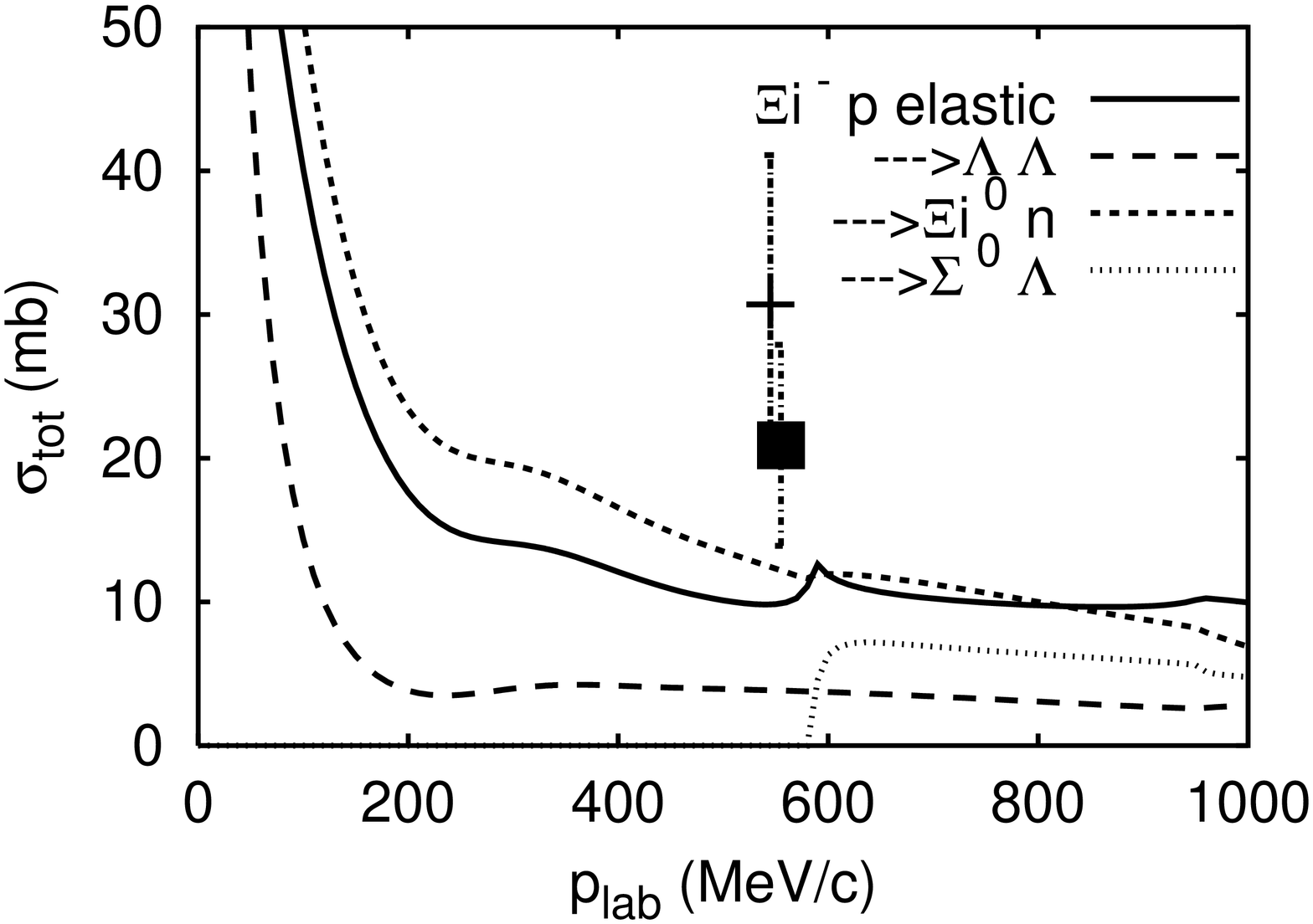}
\includegraphics[angle=-90,width=\textwidth]{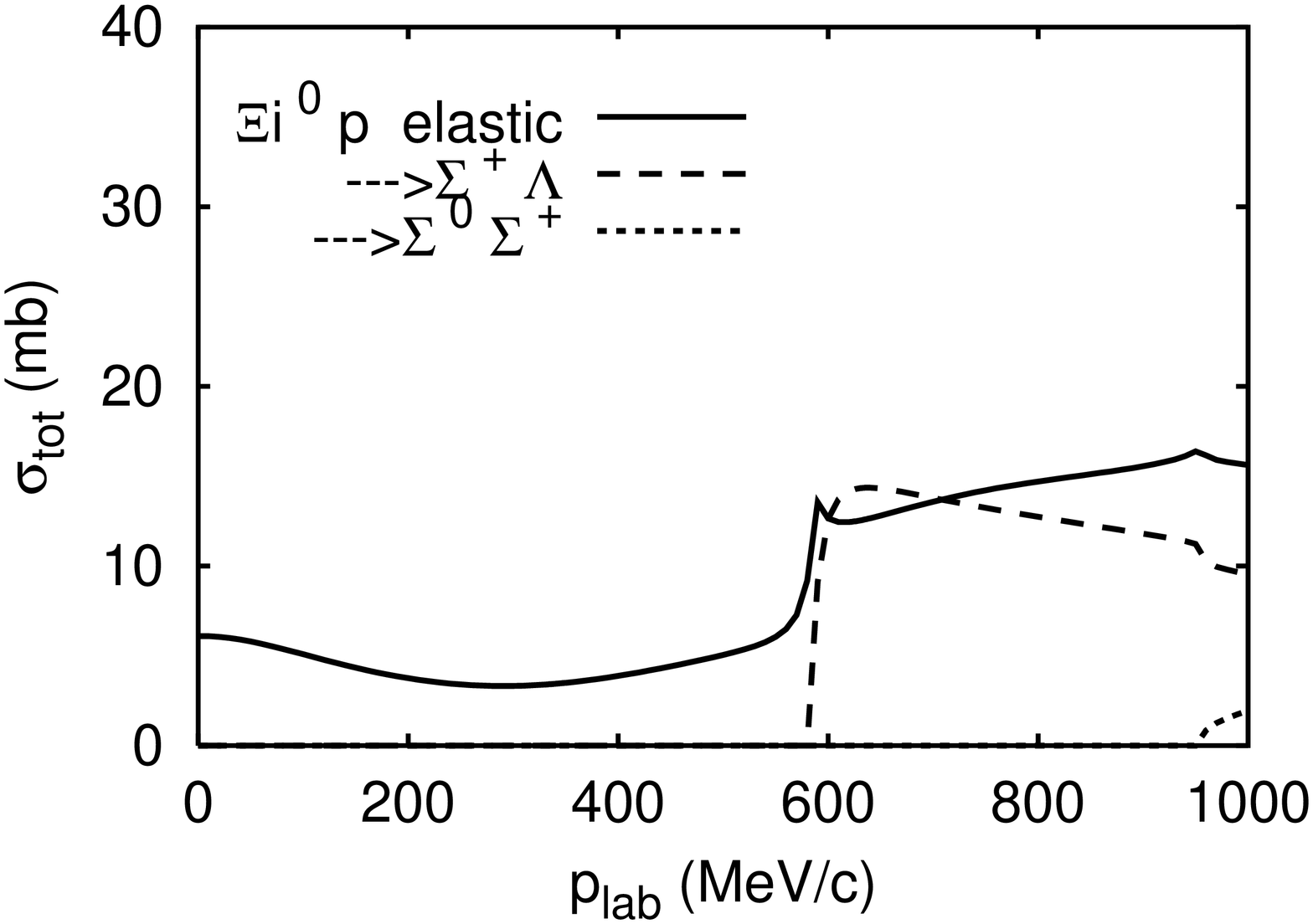}
\end{minipage}
\caption{$\Xi^- p$ and $\Xi^0 p$ total cross sections
predicted by FSS (left) and fss2 (right). For $\Xi^- p$ cross
sections both of the isospin $I=0$ and 1 channels contribute,
while for $\Xi^- n$ (or $\Xi^0 p$) only the channel
with $I=1$ contributes. The Coulomb force is neglected.}
\label{fig9}
\end{figure}

We show some comparison of the $\Xi^- N$ total cross sections
which are recently obtained
from the BNL-E906 experiment. \cite{TAM01}
The in-medium $\Xi^- N$ total cross sections
around the momentum region $p_{\rm lab} \sim 550~\hbox{MeV}/c$ are
estimated as $\sigma_{\Xi N}(\hbox{in~medium})
=30 \pm 6.7^{+3.7}_{-3.6}~\hbox{mb}$. More detailed analysis
using the Eikonal approximation \cite{YA01} gives
$\sigma_{\Xi N}(\hbox{in~medium})
=20.9 \pm 4.5^{+2.5}_{-2.4}~\hbox{mb}$.
They have also estimated the cross section ratio
$\sigma_{\Xi^- p}/\sigma_{\Xi^- n}=1.1^{+1.4}_{-0.7}
\hbox{}^{+0.7}_{-0.4}$.
If we compare these experimental data with the
FSS and fss2 predictions depicted in Fig.~\ref{fig9},
we find that the FSS predictions in the left panels
seem to be more favorable. However, we definitely need
more experimental data with higher statistics.

\begin{table}[hb]
\caption{
The relationship between the isospin basis
and the flavor-$SU_3$ basis for the $B_8 B_8$ systems. 
The flavor-$SU_3$ symmetry is given by the Elliott
notation $(\lambda \mu)$.
The heading $\CP$ denotes the flavor exchange symmetry,
$S$ the strangeness, and $I$ the isospin.}
\renewcommand{\arraystretch}{1.4}
\setlength{\tabcolsep}{5mm}
\begin{center}
\begin{tabular}{cccc}
\hline
$S$ & $B_8\,B_8~(I)$ & ${\cal P}=+1$ (symmetric)
    & ${\cal P}=-1$ (antisymmetric) \\
\cline{3-4}
    & & $\hbox{}^1 E$ \quad or \quad $\hbox{}^3 O$
    & $\hbox{}^3 E$ \quad or \quad $\hbox{}^1 O$ \\ 
\hline
$0$ & $NN~(0)$ & $-$ & $(03)$  \\
  & $NN~(1)$ & $(22)$ & $-$ \\
\hline
  & $\Lambda N~(1/2)$ &  ${1 \over \sqrt{10}}[(11)_s + 3 (22)]$ &
${1 \over \sqrt{2}} [ -(11)_a + (03) ]$ \\
$-1$ & $\Sigma N~(1/2)$ & ${1 \over \sqrt{10}}[3 (11)_s - (22)]$ &
${1 \over \sqrt{2}} [ (11)_a + (03) ]$ \\
  & $\Sigma N~(3/2)$ & $(22)$ & $(30)$ \\[1mm]
\hline
  &  $\Lambda\Lambda~(0)$ & $\frac{1}{\sqrt{5}}(11)_s
+\frac{9}{2\sqrt{30}}(22)+\frac{1}{2\sqrt{2}}(00)$ & $-$ \\
  & $\Xi N~(0)$ & $\frac{1}{\sqrt{5}}(11)_s - \sqrt{\frac{3}{10}}(22)
 +\frac{1}{\sqrt{2}}(00)$ & $(11)_a$ \\
  & $\Xi N~(1)$ & $\sqrt{\frac{3}{5}}(11)_s+\sqrt{\frac{2}{5}}(22)$
  & $\frac{1}{\sqrt{3}}[-(11)_a+(30)+(03) ]$ \\
$-2$ & $\Sigma\Lambda~(1)$ & $-\sqrt{\frac{2}{5}}(11)_s
+\sqrt{\frac{3}{5}}(22)$ & $\frac{1}{\sqrt{2}}[(30)-(03)]$ \\
  & $\Sigma\Sigma~(0)$ & $\sqrt{\frac{3}{5}}(11)_s
-\frac{1}{2\sqrt{10}}(22)-\sqrt{\frac{3}{8}}(00)$ & $-$ \\
  & $\Sigma\Sigma~(1)$ & $-$ & $\frac{1}{\sqrt{6}}
[2(11)_a+(30)+(03)]$ \\
  & $\Sigma\Sigma~(2)$ & $(22)$ & $-$ \\[1mm]
\hline
  & $\Xi \Lambda~(1/2)$ &  ${1 \over \sqrt{10}}[(11)_s + 3 (22)]$ &
${1 \over \sqrt{2}} [ -(11)_a + (30) ]$ \\
$-3$ & $\Xi \Sigma~(1/2)$ & ${1 \over \sqrt{10}} [3(11)_s-(22)]$ &
${1 \over \sqrt{2}} [ (11)_a + (30) ]$ \\
   & $\Xi \Sigma~(3/2)$ & $(22)$ & $(03)$ \\[1mm]
\hline
$-4$ & $\Xi \Xi~(0)$ & $-$ & $(30)$  \\
   & $\Xi \Xi~(1)$ & $(22)$ & $-$ \\
\hline
\end{tabular}
\end{center}
\label{table2}
\end{table}

\subsection{Characteristics of the $B_8 B_8$ Interactions by fss2}

Since our quark-model parameters are fixed by using the experimental
data in the $NN$ and $YN$ sectors, the $B_8 B_8$ interactions
beyond the strangeness $S=-1$ are all model predictions. \cite{FU01b}
For the systematic understanding of these interactions, it is
convenient to discuss them by using the $SU_3$ representation basis
for the two-baryon systems. This is because the quark-model
Hamiltonian is approximately $SU_3$ scalar,
and the interactions
with the same $SU_3$ label $(\lambda \mu)$ should
have very similar characteristics,
as long as the flavor symmetry breaking is negligible.
Table~\ref{table2} illustrates how the two baryon systems
in the isospin basis are classified
as the superposition of the $SU_3$ basis.
It has entirely different structure between
the flavor symmetric and antisymmetric cases.
In the $\hbox{}^1S_0$ state, for example,
there appear many states having the dominant (22) components.
The $S$-state $B_8 B_8$ interactions for these states should
be very similar to that for the $NN$ $\hbox{}^1S_0$ state.
The $(11)_s$ component is completely Pauli forbidden
and is characterized by the strong repulsion
originating from the quark Pauli principle.
The (00) component in the H-particle channel
is attractive from the color-magnetic interaction.
On the other hand, in the $NN$ sector the $\hbox{}^3S_1$ state
with $I=0$ is composed of the pure (03) state,
and the deuteron is bound in this channel
owing to the strong one-pion tensor force.
This $SU_3$ state in the flavor antisymmetric case
is converted to the (30) state
in the larger side of the strangeness.
Since the (30) state is almost Pauli forbidden,
the interaction is strongly repulsive.
Therefore, the $\Xi \Xi$ interaction is not so attractive
as $NN$, since they are combinations of (22) and (30).
The other $SU_3$ state $(11)_a$ turns out
to have very weak interaction.
After all, the strangeness $S=-2$ sector is most difficult,
since it is a turning point of the strangeness.  
It is also interesting to see that $\Xi \Sigma$ channel
with $I=3/2$ should be fairly attractive,
since the same (22) and (03) $SU_3$ states as in the $NN$ system
appear in this isospin channel.

\begin{figure}[t]
\begin{minipage}[h]{0.48\textwidth}
\vspace{-9mm}
\centerline{\epsfxsize=\textwidth\epsfbox{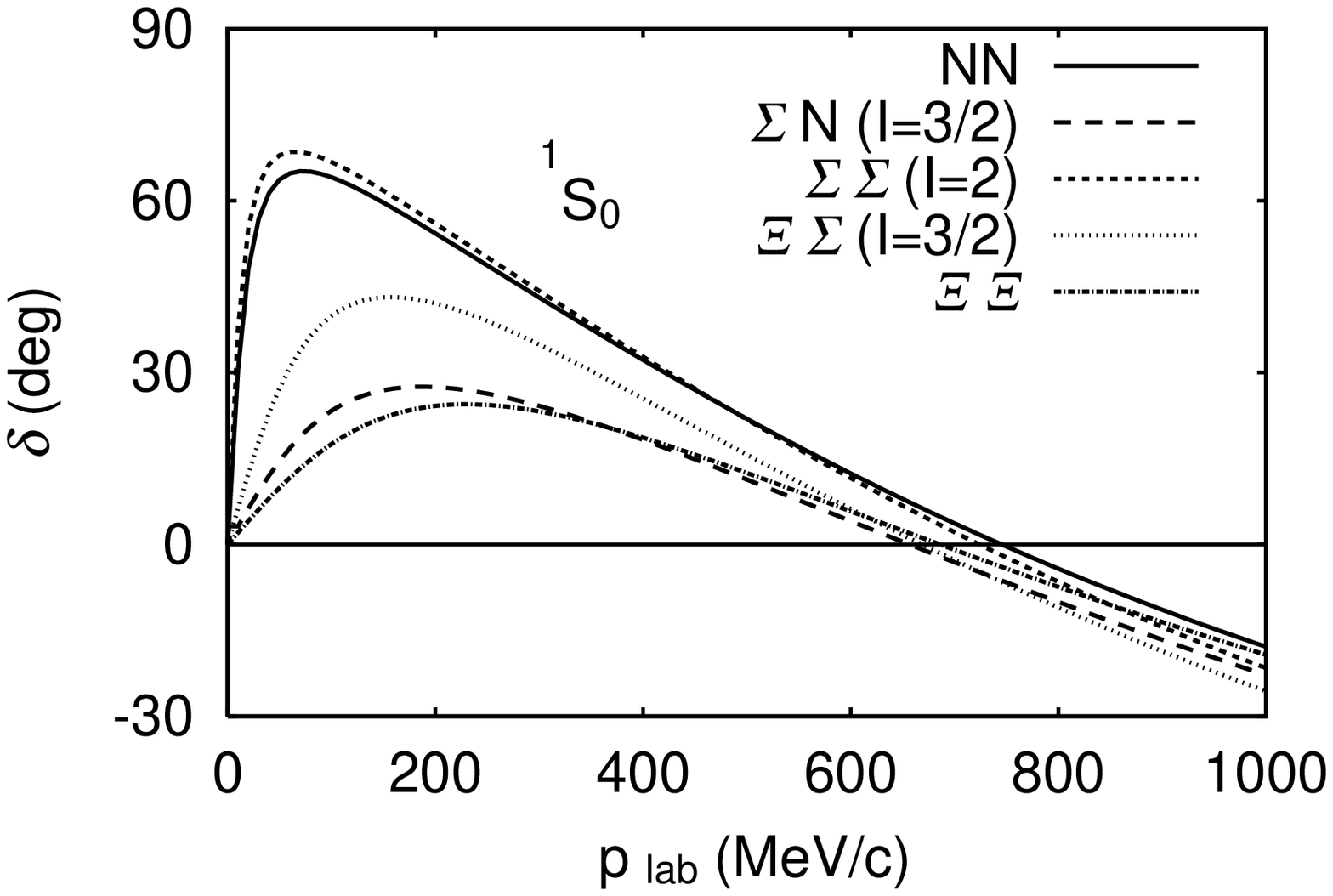}}
\caption{
$\hbox{}^1S_0$ phase shifts for various $B_8 B_8$ interactions
with the pure (22) state. The model is fss2.}
\label{fig10}
\end{minipage}
\hfill
\begin{minipage}[h]{0.48\textwidth}
\includegraphics[angle=-90,width=\textwidth]{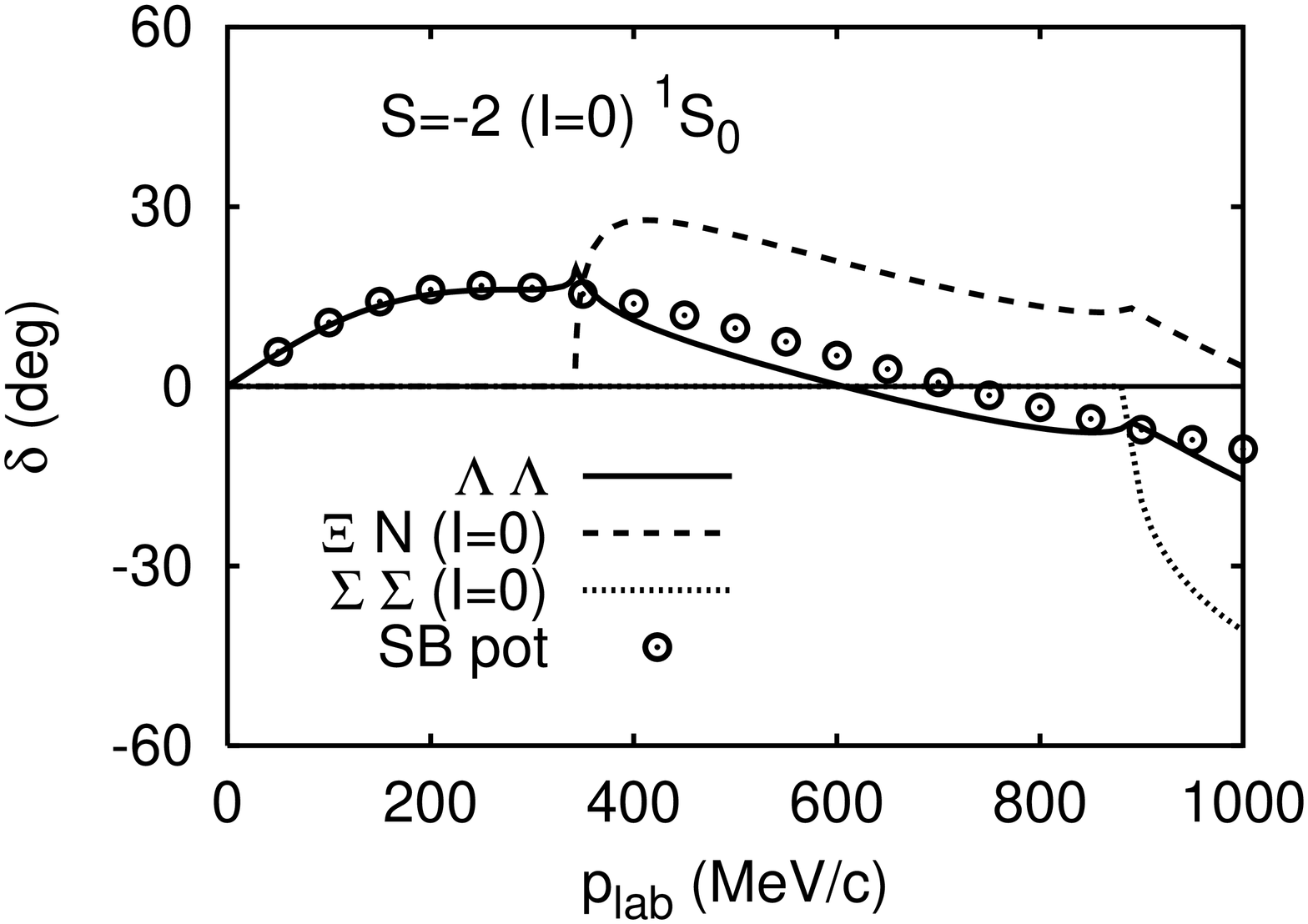}
\caption{
$\hbox{}^1S_0$ phase shifts, predicted by fss2,
in the $\Lambda \Lambda$-$\Xi N$-$\Sigma \Sigma$ coupled-channel
system with the isospin $I=0$.
The single-channel phase shift of the $\Lambda \Lambda$ channel,
predicted by the SB potential, is also shown in circles.
}
\label{fig11}
\end{minipage}
\end{figure}

Figure \ref{fig10} shows fss2 predictions
of the $\hbox{}^1S_0$ phase shifts
for various $B_8 B_8$ interactions
having the pure (22) configuration.
Although the $\Sigma \Sigma$ interaction
with the isospin $I=2$ is very similar
to the $NN$ interaction, the other interactions generally
get weaker as the strangeness involved becomes larger.
This is a combined effect of the flavor symmetry breaking
in the quark and meson-exchange contributions.
In particular, the $\Xi \Xi$ interaction has the lowest
rise of the phase shift, which is less than $30^\circ$.
Accordingly, the the $\Xi \Xi$ total cross sections
become much smaller than the other systems.
We find that the $NN$ interaction is the
strongest and has the largest cross sections
among any combinations of the octet baryons.

Figure \ref{fig11} shows the $\hbox{}^1S_0$ phase shift curves,
predicted by fss2,
in the $\Lambda \Lambda$-$\Xi N$- $\Sigma \Sigma$ coupled-channel
system with the isospin $I=0$.
The maximum peak of the $\Lambda \Lambda$ phase shift is
less than $20^\circ$, which is much smaller than the previous
predictions by various models.
This result is in good agreement with the recent experimental data
for the double $\Lambda$ hypernucleus
$\hbox{}^{\ 6}_{\Lambda \Lambda}\hbox{He}$.
The finding of this event, called the Nagara
event, \cite{TA01} is one of the most important contributions
in recent years, since the assignment of the decaying scheme
is very definite.
The $\Delta B_{\Lambda \Lambda}$ value, 
defined by $\Delta B_{\Lambda \Lambda}=B_{\Lambda \Lambda}
(\hbox{}^{\ \,6}_{\Lambda \Lambda}\hbox{He})
-2B_{\Lambda}(\hbox{}^{\,5}_{\Lambda}\hbox{He})$,
is deduced from this event as $\Delta B_{\Lambda \Lambda}
=1.01 \pm 0.20$ MeV, which implies a weak attraction
for the $\Lambda \Lambda$ interaction.
Our $G$-matrix calculation of this system
yields an almost right answer, $\Delta B_{\Lambda \Lambda}
=1.12 \sim 1.24$ MeV, by taking into account the important
contribution from the $\alpha$-particle rearrangement
energy. \cite{KO03} In the next section, we will solve the
Faddeev equation of the $\Lambda \Lambda \alpha$ system,
by using the $\Lambda \Lambda$ $T$-matrix predicted by
our QM interaction.

Some of the following characteristics
of fss2 for the $B_8 B_8$ interactions
are very much different from the Nijmegen predictions
given by Stoks and Rijken. \cite{ST99}

\medskip

\begin{enumerate}
\item[$\circ$] There is no bound state in the $B_8 B_8$ system,
except for the deuteron.
\item[$\circ$] The $\Xi \Xi$ total cross sections
are not so large as the $NN$ total cross sections.
\item[$\circ$] The $\Xi N$ interaction has a strong
isospin dependence like the $\Sigma N$ interaction.
\item[$\circ$] The $\Xi^- \Sigma^-$ (namely,
$\Xi \Sigma (I=3/2)$) interaction is fairly attractive.
\end{enumerate}

\section{Faddeev calculation}

\subsection{Three-nucleon bound state}

Since our QM $B_8 B_8$ interaction describes
the short-range repulsion very differently
from the meson-exchange potentials, it is interesting
to examine the three-nucleon system predicted by fss2 and FSS.
Here we solve the Faddeev equation
for $\hbox{}^3\hbox{H}$, by directly
using the QM RGM kernel in the isospin basis. \cite{TR02}
Our Faddeev calculation is the full 50-channel calculation up to
the maximum angular momentum $J=6$,
and the values are almost completely
converged as seen in Table~\ref{table3}. \cite{PANIC02}
The fss2 prediction, $-8.52$ MeV, seems to be overbound
in comparison with the experimental
value $E^{\rm exp}(\hbox{}^3\hbox{H})=-8.48$ MeV.
In fact, this is not the case,
since all these calculations neglect
the charge dependence of the $NN$ interaction, in which
the $\hbox{}^1S_0$ interaction of the $nn$ system
is less attractive than that of the $np$ system.
The effect of the charge dependence is estimated
to be $-0.19$ MeV for the triton binding energy. \cite{MA89}
If we take this into account,
our result is still 150 keV less bound.
If we attribute this difference to the effect
of the three-nucleon force, it is by far smaller than the
generally accepted values, $0.5 \sim 1$ MeV \cite{No00},
predicted by many Faddeev calculations employing modern
realistic meson-theoretical $NN$ interactions.
The charge root-mean-square radii
of $\hbox{}^3\hbox{H}$ and $\hbox{}^3\hbox{He}$ are
correctly reproduced.
An important point here is that
we can reproduce enough binding energy of the triton,
without reducing the deuteron $D$-state probability.

\begin{table}[t]
\caption{
The three-nucleon bound state properties predicted by the Faddeev
calculation with fss2 and FSS. The $np$ interaction is used
in the isospin basis. 
The heading ``No. of channels'' implies the number of two-nucleon
channels included; $n_{\rm max}$ is the dimension
of the diagonalization for the Faddeev equation;
$\varepsilon_{NN}$ is the $NN$ expectation value
determined self-consistently;
$E(\hbox{}^3\hbox{H})$ is the ground-state energy; and 
$\protect\sqrt{\langle r^2\rangle_{\hbox{}^3{\rm H}}}$ 
($\protect\sqrt{\langle r^2\rangle_{\hbox{}^3{\rm He}}}$) is
the charge rms radius for $\hbox{}^3{\rm H}$ ($\hbox{}^3{\rm He}$),
including the proton and neutron size corrections.
The Coulomb force and the relativistic corrections are neglected.}
\renewcommand{\arraystretch}{1.2}
\setlength{\tabcolsep}{4mm}
\begin{center}
\begin{tabular}{ccccccc}
\hline
model & No. of & $n_{\rm max}$ & $\varepsilon_{NN}$
 & $E(\hbox{}^3\hbox{H})$
 & $\protect\sqrt{\langle r^2\rangle_{\hbox{}^3{\rm H}}}$
 & $\protect\sqrt{\langle r^2\rangle_{\hbox{}^3{\rm He}}}$ \\
 & channels & & (MeV) & (MeV)  & (fm) & (fm) \\
\hline
     &  2 ($S$)       &   2,100 & 2.361 & $-7.807$ & 1.80 & 1.96 \\
     &  5 ($SD$)      &   5,250 & 4.341 & $-8.189$ & 1.75 & 1.92 \\
fss2 & 10 ($J \le 1$) &  10,500 & 4.249 & $-8.017$ & 1.76 & 1.94 \\
     & 18 ($J \le 2$) &  18,900 & 4.460 & $-8.439$ & 1.72 & 1.90 \\
     & 34 ($J \le 4$) &  35,700 & 4.488 & $-8.514$ & 1.72 & 1.90 \\
     & 50 ($J \le 6$) & 112,500 & 4.492 & $-8.519$ & 1.72 & 1.90 \\
\hline
     &  2 ($S$)       &   2,100 & 2.038 & $-7.675$ & 1.83 & 1.99 \\
     &  5 ($SD$)      &   5,250 & 3.999 & $-8.034$ & 1.78 & 1.95 \\
FSS  & 10 ($J \le 1$) &  10,500 & 3.934 & $-7.909$ & 1.78 & 1.97 \\
     & 18 ($J \le 2$) &  18,900 & 4.160 & $-8.342$ & 1.74 & 1.93 \\
     & 34 ($J \le 4$) &  35,700 & 4.175 & $-8.390$ & 1.74 & 1.92 \\
     & 50 ($J \le 6$) & 112,500 & 4.177 & $-8.394$ & 1.74 & 1.92 \\
\hline
\end{tabular}
\end{center}
\label{table3}
\end{table}

The self-consistent energy of the two-cluster RGM kernel,
$\varepsilon_{NN}$, has a clear physical meaning
related to the decomposition of the total
triton energy $E$ into the kinetic-energy and
potential-energy contributions:
$\varepsilon_{NN}=E/3+\langle H_0 \rangle/6$.
Table \ref{table4} shows this decomposition,
together with the results
of CD-Bonn and AV18 potentials. \cite{No00}
We find that our quark model results by FSS and fss2 are just
middle between these potentials, which have very different
strengths of the tensor force.

\begin{table}[b]
\caption{
Decomposition of the total triton energy $E$ into
the kinetic-energy and potential-energy contributions:
$E=\langle H_0 \rangle+\langle V \rangle$.
The unit is in MeV. In the present framework,
this is given by the expectation
value $\varepsilon_{NN}$ of the two-cluster Hamiltonian
with respect to the Faddeev solution, which is determined
self-consistently. The results of CD-Bonn and AV18 are
taken from Ref.~\protect\citen{No00}.
}
\renewcommand{\arraystretch}{1.2}
\setlength{\tabcolsep}{9mm}
\begin{center}
\begin{tabular}{ccccc}
\hline
model & $\varepsilon_{NN}$ & $E$ & $\langle H_0\rangle$
      & $\langle V \rangle$ \\
\hline
fss2     &  4.492 & $-8.519$ & 43.99 & $-52.51$ \\
FSS      &  4.177 & $-8.394$ & 41.85 & $-50.25$ \\
CD-Bonn  &  3.566 & $-8.012$ & 37.42 & $-45.43$ \\
AV18     &  5.247 & $-7.623$ & 46.73 & $-54.35$ \\
\hline
\end{tabular}
\label{table4}
\end{center}
\end{table}

\subsection{The hypertriton}

Next, we apply our QM $NN$ and $YN$ interactions
to the hypertriton ($\hbox{}^3_\Lambda \hbox{H}$) with
the small separation energy of the $\Lambda$-particle,
${B_\Lambda}^{\rm exp}=130 \pm 50~\hbox{keV}$. \cite{hypt}
Since the $\Lambda$-particle is far apart from
the two-nucleon subsystem, the on-shell properties
of the $\Lambda N$ and $\Sigma N$ interactions are expected to be
well reflected in this system. 
In particular, this system is very useful to learn the relative
strength of $\hbox{}^1S_0$ and $\hbox{}^3S_1$ attractions
of the $\Lambda N$ interaction, since the $\hbox{}^1S_0$ component
plays a more important role than $\hbox{}^3S_1$ in this system
and the available low-energy $\Lambda p$ total cross section
data cannot discriminate many possible combinations
of the $\hbox{}^1S_0$ and $\hbox{}^3S_1$ interactions.
In fact, Ref.~\citen{MI95,MI00,No02} showed that
most meson-theoretical interactions fail to bind the hypertriton,
except for the Nijmegen soft-core potentials NSC89, \cite{NSC89}
NSC97f and NSC97e. \cite{NSC97} It is also pointed out 
in Refs.~\citen{MI95} and \citen{No02} that a small admixture
of the $\Sigma NN$ components less than 1\% is
very important for this binding.
We therefore carry out the $\Lambda NN$-$\Sigma NN$ coupled-channel
Faddeev calculation, by properly taking into account the existence
of the $SU_3$ Pauli forbidden state $(11)_s$ at the quark level.

\begin{table}[t]
\caption{Results of the hypertriton Faddeev
calculations by fss2 and FSS.
The heading $E$ is the hypertriton
energy measured from the $\Lambda NN$ threshold; $B_\Lambda$ is
the $\Lambda$ separation energy; $\varepsilon_{NN}$
($\varepsilon_{\Lambda N}$) is the $NN$ ($\Lambda N$) expectation
value determined self-consistently;
and $P_\Sigma$ is the $\Sigma NN$ probability in percent.
The norm of admixed redundant components is less than $10^{-9}$.}
\renewcommand{\arraystretch}{1.2}
\setlength{\tabcolsep}{5mm}
\begin{center}
\begin{tabular}{@{}ccccccc}
\hline
model & No. of & $E$ & $B_\Lambda$
& $\varepsilon_{NN}$ & $\varepsilon_{\Lambda N}$
& $P_\Sigma$ \\
      & channels & (MeV) & (keV) & (MeV) & (MeV) & $(\%)$ \\
\hline
 &   6 ($S$)  & $-2.362$ & 137 & $-1.815$ & 3.548 & 0.450 \\
 &  15 ($SD$) & $-2.423$ & 198 & $-1.762$ & 5.729 & 0.652 \\
fss2 &  30 ($J \le 1$) & $-2.403$ & 178 & $-1.786$ & 5.664 & 0.615 \\
 & 54 ($J \le 2$) & $-2.498$ & 273 & $-1.673$ & 5.974 & 0.777 \\
 & 102 ($J \le 4$) & $-2.513$ & 288 & $-1.658$ & 6.022 & 0.804 \\
 & 150 ($J \le 6$) & $-2.514$ & 289 & $-1.657$ & 6.024 & 0.805 \\
\hline
 &   6 ($S$)  & $-2.910$ & 653 & $-1.309$ & 3.984 & 1.022 \\
 &  15 ($SD$) & $-2.967$ & 710 & $-1.433$ & 6.171 & 1.200 \\
FSS &  30 ($J \le 1$) & $-2.947$ & 691 & $-1.427$ & 6.143 & 1.191 \\
 & 54 ($J \le 2$) & $-3.121$ & 865 & $-1.323$ & 6.467 & 1.348 \\
 & 102 ($J \le 4$) & $-3.134$ & 877 & $-1.317$ & 6.488 & 1.360 \\
 & 150 ($J \le 6$) & $-3.134$ & 878 & $-1.317$ & 6.488 & 1.361 \\
\hline
\end{tabular}
\end{center}
\label{table5}
\end{table}

Table \ref{table5} shows the results of the Faddeev calculations
using fss2 and our previous model FSS.
In the 15-channel calculation including the $S$ and $D$ waves
of the $NN$ and $YN$ interactions,
we have already obtained $B_\Lambda=-\varepsilon_d
-E(\hbox{}^3_\Lambda \hbox{H}) \sim 200$ keV for fss2.
The convergence with the extension to the higher partial waves
is very rapid, and the total angular-momentum
of the baryon pairs with $J \leq 4$ is
good enough for 1 keV accuracy.
As for the converged $B_\Lambda$ values
with 150-channel $\Lambda NN$ and $\Sigma NN$ configurations,
we obtain $B_\Lambda=289$ keV and
the $\Sigma NN$ component $P_\Sigma=0.80\,\%$ for the
fss2 prediction,
and $B_\Lambda=878$ keV and $P_\Sigma=1.36\,\%$ for FSS.

Table \ref{table6} shows the correlation
between the $\Lambda$ separation
energy $B_\Lambda$ and the $\hbox{}^1S_0$ and $\hbox{}^3S_1$
effective range parameters of FSS,
fss2 and NSC89 $\Lambda N$ interactions.
Although all of these $\Lambda N$ interactions reproduce
the low-energy $\Lambda N$ total cross section data
below $p_\Lambda \sim 300~\hbox{MeV}/c$ within the experimental
error bars, our quark-model interactions seem to be slightly more
attractive than the Nijmegen soft-core potential NSC89 \cite{NSC89}.
The model FSS gives a large overbinding
since the $\hbox{}^1S_0$ $\Lambda N$ interaction is strongly
attractive. The phase-shift difference
of the $\hbox{}^1S_0$ and $\hbox{}^3S_1$ states
at $p_\Lambda \sim 200~\hbox{MeV}/c$ is $\delta(\hbox{}^1S_0)
-\delta(\hbox{}^3S_1) \sim 29^\circ$ for FSS,
while $\sim 7^\circ$ for fss2. Since the present fss2 result
is still slightly overbound, this difference should be somewhat
smaller in order to reproduce the correct experimental
value ${B_\Lambda}^{\rm exp}=130 \pm 50~\hbox{keV}$.
From the two results given by fss2 and FSS, we can extrapolate
the desired difference to be $0^\circ \sim 2^\circ$,
which is consistent with the result in Ref.~\citen{NE02} using
simulated interactions of the Nijmegen models.

\begin{table}[t]
\caption{$\hbox{}^1S_0$ and $\hbox{}^3S_1$ effective
range parameters of various $\Lambda N$ interactions
and the $\Lambda$ separation energy $B_\Lambda$ of the
hypertriton. The $B_\Lambda$ value for NSC89 is taken
from Ref.~\protect\citen{No02}.
}
\renewcommand{\arraystretch}{1.2}
\setlength{\tabcolsep}{5mm}
\begin{center}
\begin{tabular}{@{}cccccc}
\hline
model & $a_s$ (fm) & $r_s$ (fm) & $a_t$ (fm) & $r_t$ (fm)
& $B_\Lambda$ (keV) \\
\hline
FSS \protect\cite{FU96b} & $-5.41$ & 2.26 & $-1.03$ & 4.20
& 878 \\
fss2 \protect\cite{FU02a,FU01b} & $-2.59$ & 2.83 & $-1.60$ & 3.01
& 289 \\
NSC89 \protect\cite{NSC89} & $-2.78$ & 2.88 & $-1.41$ & 3.11
& 143 \\
\hline
\end{tabular}
\label{table6}
\end{center}
\end{table}

\begin{table}[b]
\caption{Decomposition of the $NN$ and $YN$ expectation
values ($\varepsilon_{NN}$ and $\varepsilon_{YN}$), the
deuteron energy ($-\varepsilon_d$) and the total three-body
energy $E$ to the kinetic-energy and potential-energy
contributions. The unit is in MeV. 
The results for NSC89 are taken from Ref.~\protect\citen{MI95}.}
\renewcommand{\arraystretch}{1.4}
\setlength{\tabcolsep}{10mm}
\begin{center}
\begin{tabular}{@{}cc@{}cc}
\hline
model & $h_{NN}+V_{NN}=\varepsilon_{NN}$ &
& $h_d+V_d=-\varepsilon_d$ (deuteron) \\
\hline
FSS   & $19.986-21.303=-1.317$ & & $16.982-19.238=-2.256$ \\
fss2  & $19.376-21.032=-1.657$ & & $17.495-19.720=-2.225$ \\
NSC89 & $20.48 -22.25 =-1.77$  & & $19.304-21.528=-2.224$ \\
\hline
model & $h_{YN}+V_{YN}=\varepsilon_{YN}$ &
& $\langle H \rangle+\langle V \rangle=E$ \\
\hline
FSS   & $10.036-4.602=5.435$ & & $27.372-30.506=-3.134$ \\
fss2  &  $8.071-2.671=5.401$ & & $23.860-26.374=-2.514$ \\
NSC89 &  $7.44 -3.54 =3.90$  & & $23.45 -25.79 =-2.34$  \\
\hline
\end{tabular}
\end{center}
\label{table7}
\end{table}

Table \ref{table5} also shows that the expectation value
of the $NN$ Hamiltonian, $\varepsilon_{NN}$, determined
self-consistently is rather close
to the deuteron energy $-\varepsilon_d$, especially in fss2.
This feature is even marked if we decompose these energies
to the kinetic-energy and potential-energy contributions.
Table \ref{table7} shows this decomposition with respect
to fss2, FSS and NSC89. (For this comparison,
we use the definition of the kinetic-energy part of the deuteron,
given by $h_d=\langle \chi_d|h_{NN}|\chi_d \rangle/\langle \chi_d|
\chi_d \rangle$, where $\chi_d$ is the RGM relative wave function
between the neutron and the proton.)
In fss2, the kinetic-energy of the $NN$ subsystem
is 1.88 MeV larger than that of the deuteron,
which implies that the $NN$ subsystem
shrinks by the effect of the outer $\Lambda$-particle, 
in comparison with the deuteron in the free space.
In NSC89, this difference is even smaller; i.e., 1.18 MeV.
These results are consistent with the fact
that the hypertriton in NSC89 is more loosely
bound ($B_\Lambda=143$ keV) \cite{No02} than in fss2 (289 keV),
and the $\Lambda$-particle is very far apart from the $NN$ cluster.
The $\Sigma NN$ probability in NSC89
is $P_\Sigma=0.5\,\%$. \cite{MI95}
Table \ref{table7} also lists the kinetic-energy and
potential-energy decompositions
for the $\Lambda N$-$\Sigma N$ averaged $YN$ expectation
value $\varepsilon_{YN}$ and the total energy $E$.
The kinetic energies of $\varepsilon_{YN}$ are much smaller
than those of $\varepsilon_{NN}$, which indicates that the
relative wave functions between the hyperon and the nucleon are
widely spread in the configuration space.
The comparison of the total-energy decomposition shows that
the wave functions of fss2 and NSC89 may be very similar.
A clear difference between fss2 and NSC89 appears
in the roles of higher partial waves. The energy increase
due to the higher partial waves than the $S$ and $D$ waves
is 91 keV in fss2 and 168 keV in FSS, respectively.
On the other hand, the results
in Refs.~\citen{MI95} and \citen{No02} imply that
this is only 20 - 30 keV in the case of NSC89.
This difference can originate from both
of the $NN$ and $YN$ interactions. Since the characteristics 
of the meson-theoretical $YN$ interactions in higher partial
waves are a priori unknown, more detailed analysis of the
fss2 results might shed light on the adequacy of the
quark-model baryon-baryon interactions.

\begin{figure}[b]
\begin{minipage}[ht]{0.49\textwidth}
\includegraphics[angle=-90,width=\textwidth]{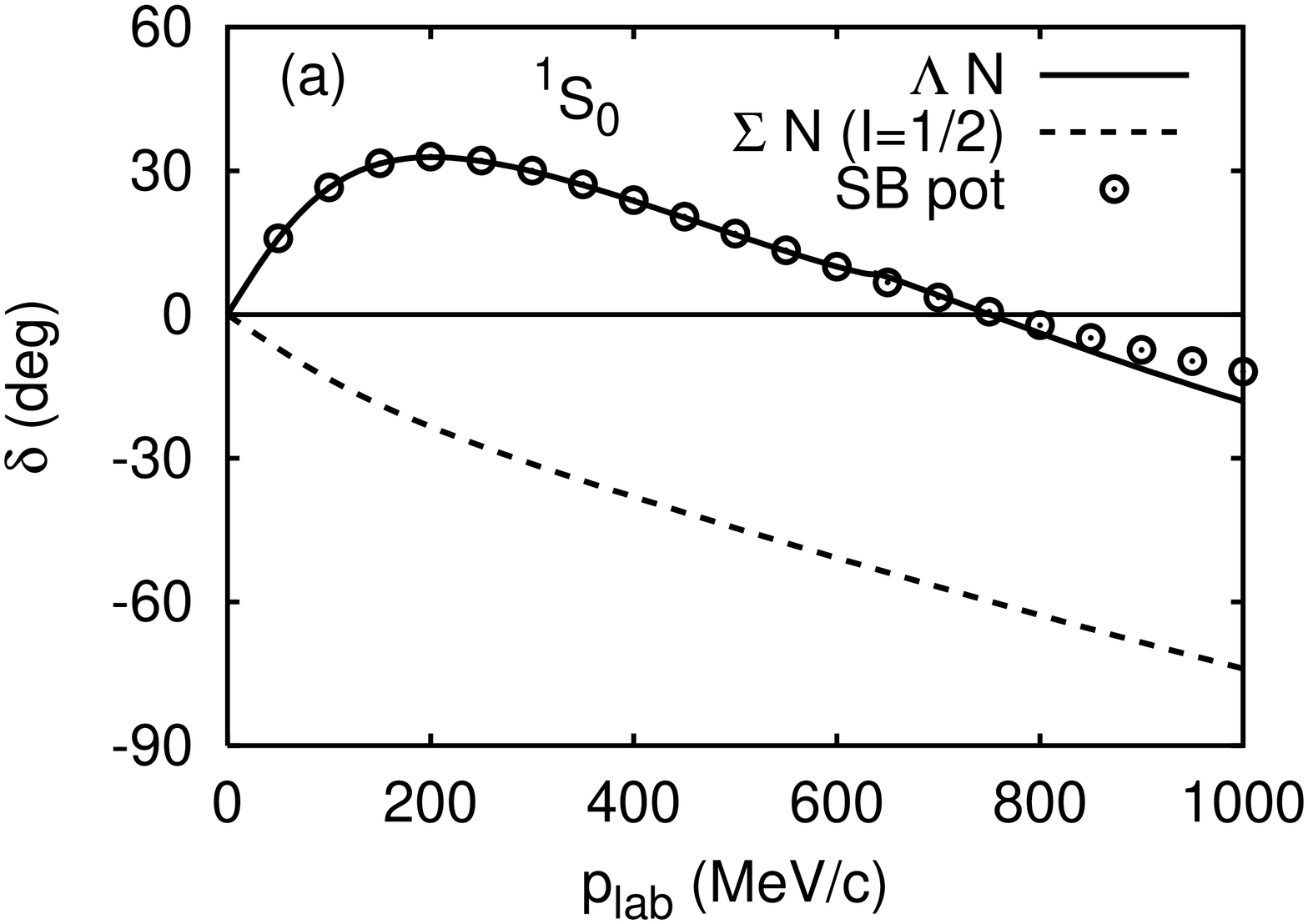}
\end{minipage}
%
\hspace{\fill}
\begin{minipage}[ht]{0.49\textwidth}
\includegraphics[angle=-90,width=\textwidth]{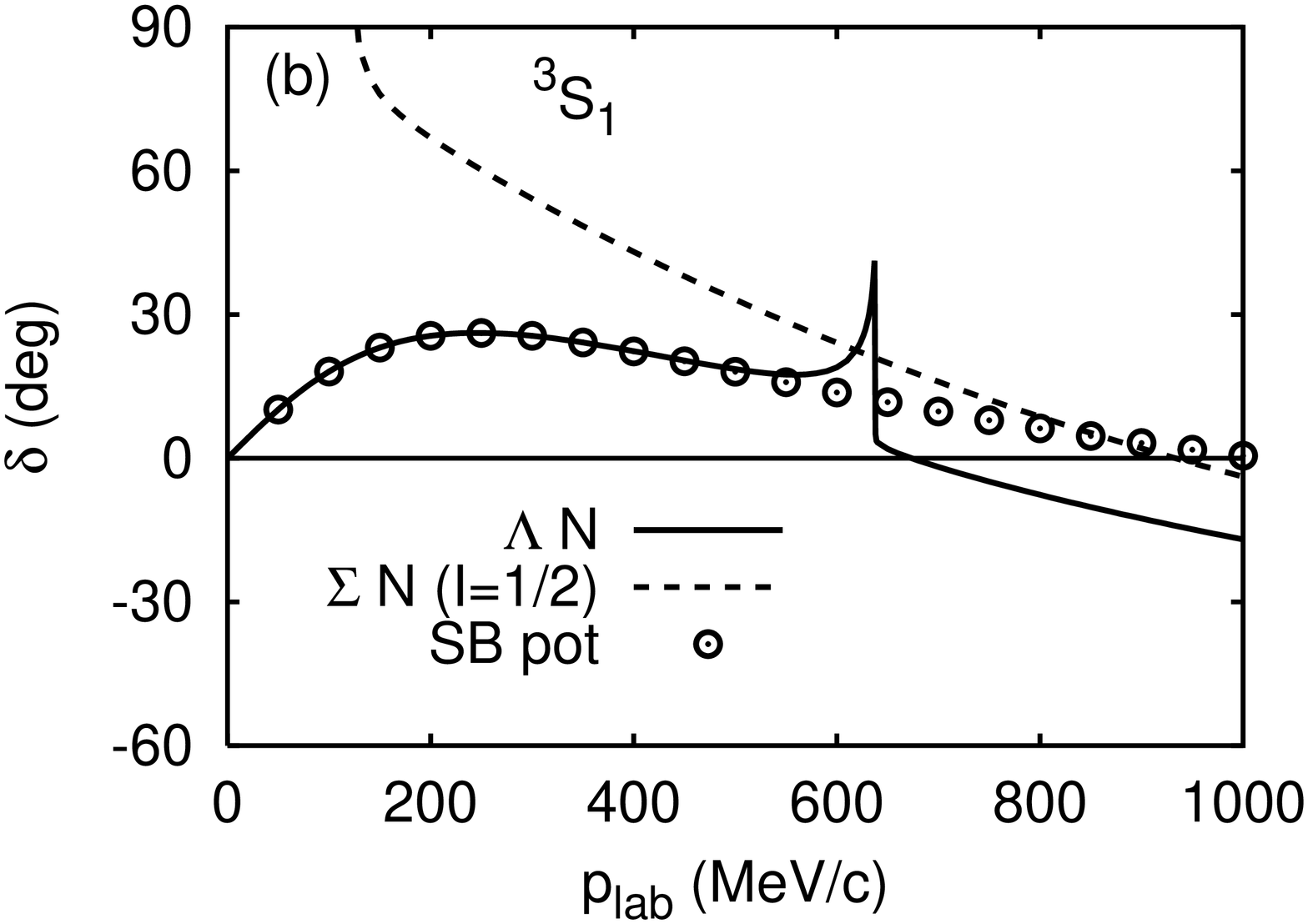}
\end{minipage}
\caption{$\Lambda N$-$\Sigma N$  $\hbox{}^1S_0$ (a)
and $\hbox{}^3S_1$ (b) phase shifts
for the isospin $I=1/2$ channel,
calculated with fss2 (solid and dashed curves)
and with the SB potential (circles).}
\label{fig12}
\end{figure}

\subsection{The $\alpha \alpha \Lambda$ system
for $\hbox{}^9_{\Lambda}\hbox{Be}$}

As another typical example of three-cluster systems composed of
two identical clusters, we apply the present formalism
to the $\alpha \alpha \Lambda$ Faddeev calculation
for $\hbox{}^9_{\Lambda}\hbox{Be}$, using the $\alpha \alpha$
RGM kernel and the $\Lambda \alpha$ folding potential generated 
from a simple $\Lambda N$ effective interaction. \cite{2al}
For the $\alpha \alpha$ RGM kernel,
we use the the three-range Minnesota
force \cite{TH77} with the Majorana exchange mixture $u=0.94687$,
and the h.o. width parameter, $\nu=0.257~\hbox{fm}^{-2}$,
assumed for the $(0s)^4$ $\alpha$-clusters.
The $\alpha \alpha$ phase shifts are nicely reproduced
in the $\alpha \alpha$ RGM calculation,
using this effective $NN$ interaction.
For the $3\alpha$ system, we find that the ground-state
energy obtained by solving the present $3\alpha$ Faddeev equation
is only 1.8 MeV higher than that of the fully
microscopic $3\alpha$ RGM calculation.
The effective $\Lambda N$ interaction,
denoted by SB (Sparenberg-Baye potential) in Table\,\ref{table8},
is constructed from
the $\hbox{}^1S_0$ and $\hbox{}^3S_1$ phase shifts predicted
by the $YN$ sector of the model fss2,
by using an inversion method based on supersymmetric
quantum mechanics. \cite{SB97}
These are simple two-range Gaussian potentials which reproduce the
low-energy behavior of the $\Lambda N$ phase shifts
in Fig.\,\ref{fig12}, obtained in the full
coupled-channel calculations:
\begin{eqnarray*}
V_{\hbox{}^1S_0}(r) & = & -128.0~\exp(-0.8908~r^2)+1015~\exp(-5.383
~r^2)\ \ , \nonumber \\
V_{\hbox{}^3S_1}(r) & = & -56.31\,f\,\exp(-0.7517~r^2)
+1072~\exp(-13.74~r^2)\ \ ,
\label{pot1}
\end{eqnarray*}
where $V(r)$ in MeV and $r$ in fm is the relative distance
between $\Lambda$ and $N$.
In the $\hbox{}^3S_1$ state, the phase-shift behavior only
in the low-momentum region with $p_{\rm lab} < 600~\hbox{MeV}/c$ is
fitted, since the cusp structure is never reproduced
in the single-channel calculation.
Since any central and single-channel effective $\Lambda N$ force 
leads to the well-known overbinding
problem of $\hbox{}^5_{\Lambda}\hbox{He}$ by about 2 MeV (in the
present case, it is 1.63 MeV), \cite{DA72} the attractive part
of the $\hbox{}^3S_1$ $\Lambda N$ potential is modified to reproduce
the correct binding energy, $E^{\rm exp}(\hbox{}^5_\Lambda \hbox{He})
=-3.12\pm 0.02$ MeV, with an adjustable parameter $f=0.8923$.
This overbinding problem is mainly attributed
to the Brueckner rearrangement effect of
the $\alpha$-cluster, originating from the starting energy
dependence of the bare two-nucleon interaction
due to the addition of an extra $\Lambda$ particle. \cite{BA80} 
The odd-state $\Lambda N$ interaction is assumed to be zero.
All partial waves up to $\lambda_{\rm Max}={\ell_1}_{\rm Max}
=6$ for the $\alpha \alpha$ and $\Lambda \alpha$ pairs are included.
The direct and exchange Coulomb kernel between
the two $\alpha$-clusters is introduced at the nucleon level
with the cut-off radius, $R_C=14~\hbox{fm}$.
Table\,\ref{table8} shows the ground-state ($0^+$) and
the $2^+$ excitation energies of $\hbox{}^9_{\Lambda}\hbox{Be}$,
predicted by the SB and the other various $\Lambda N$ potentials
used by Hiyama {\em et al.} \cite{HI97}
In the present calculations using only the central force,
the SB potential with the pure Serber character can reproduce
the the ground-state and the excitation energies
within 100 - 200 keV accuracy.

\begin{table}[t]
\caption{
The ground-state energy $E_{\rm gr}(0^+)$ and
$2^+$ excitation energy $E_{\rm x}(2^+)$ in MeV,
calculated by solving the Faddeev equation
for the $\alpha \alpha \Lambda$ system
in the $LS$ coupling scheme.
The $\alpha \alpha$ RGM kernel is generated
from the three-range Minnesota force.
The $\Lambda N$ force, SB, stands for the Sparenberg-Baye
potential and NS - JB are the $G$-matrix based effective forces
used by Hiyama {\em et al.} \protect\cite{HI97}
In Ref.~\protect\citen{HI97}, the orthogonality condition
model (OCM) is used for the $\alpha \alpha$ interaction. 
The experimental values are cited from Ref.~\protect\citen{AK02}.
}
\renewcommand{\arraystretch}{1.2}
\setlength{\tabcolsep}{10mm}
\begin{center}
\begin{tabular}{cccc}
\hline
$V_{\Lambda N}$ & \multicolumn{2}{c}{$E_{\rm gr}(0^+)$ (MeV)}
  & $E_{\rm x}(2^+)$ (MeV) \\
  & present & Ref.~\protect\citen{HI97} & \\
\hline
SB & $-6.837$  &  $-$    & 2.915 \\
NS & $-6.742$  & $-6.81$ & 2.916 \\
ND & $-7.483$  & $-7.57$ & 2.935 \\
NF & $-6.906$  & $-7.00$ & 2.930 \\
JA & $-6.677$  & $-6.76$ & 2.919 \\
JB & $-6.474$  & $-6.55$ & 2.911 \\
\hline
Exp't & \multicolumn{2}{c}{$-6.62 \pm 0.04$} & 3.029(3) \\
      & & & 3.060(3) \\
\hline
\end{tabular}
\label{table8}
\end{center}
\end{table}

\subsection{The $\Lambda \Lambda \alpha$ system 
for $\hbox{}^{\ 6}_{\Lambda \Lambda} \hbox{He}$}

We can use the $\Lambda \alpha$ $T$-matrix, used in
the $\alpha \alpha \Lambda$ Faddeev calculation, to calculate
the ground-state energy
of $\hbox{}^{\ 6}_{\Lambda \Lambda} \hbox{He}$.
The full coupled-channel $T$-matrices of fss2 and FSS
with the strangeness $S=-2$ and the isospin $I=0$ are
employed for the $\Lambda \Lambda$ RGM $T$-matrix. 
Table~\ref{table9} shows the $\Delta B_{\Lambda \Lambda}$ values
in MeV, predicted by various combinations
of the $\Lambda N$ and $\Lambda \Lambda$ interactions.
The results of a simple three-range Gaussian potential,
$V_{\Lambda \Lambda}$(Hiyama), used in Ref.~\citen{HI97}
are also shown.
We find that this $\Lambda \Lambda$ potential and
our Faddeev calculation using the FSS $T$-matrix yield
very similar results with the
large $\Delta B_{\Lambda \Lambda}$ values about 3.6 MeV,
since the $\Lambda \Lambda$ phases shifts
predicted by these interactions
increase up to about $40^\circ$. The improved quark model fss2
yields $\Delta B_{\Lambda \Lambda}=1.41$ MeV.
(If we use the $\Lambda \Lambda$ single-channel $T$-matrix,
this number is reduced to 1.14 MeV.)
In Table\,\ref{table9}, results are also shown
for $V_{\Lambda \Lambda}$(SB), which is a two-range
Gaussian potential generated
from the fss2 $\hbox{}^1S_0$ $\Lambda \Lambda$ phase
shift (see Fig.~\ref{fig11}) in the full-channel calculation,
using the supersymmetric inversion method. \cite{SB97}
This potential is explicitly given by
\begin{eqnarray*}
V_{\Lambda \Lambda}({\rm SB})=-103.9~\exp(-1.176~r^2)
+658.2~\exp(-5.936~r^2)
\ \ ,
\label{pot2}
\end{eqnarray*}
where $V_{\Lambda \Lambda}({\rm SB})$ in MeV and $r$ in fm
is the relative distance between two $\Lambda$'s.
We think that the 0.5 MeV difference between our fss2 result
and the $V_{\Lambda \Lambda}$(SB) result is probably because
we neglected the full coupled-channel effects
of the $\Lambda \Lambda \alpha$ channel
to the $\Xi N \alpha$ and $\Sigma \Sigma \alpha$ channels.
We should also keep in mind that in all of these
three-cluster calculations
the Brueckner rearrangement effect of the $\alpha$-cluster
with the magnitude of about $-1$ MeV (repulsive)
is very important. \cite{KO03}
It is also reported in Ref.~\citen{SU99} that
the quark Pauli effect between the $\alpha$ cluster
and the $\Lambda$ hyperon yields a non-negligible
repulsive contribution of 0.1 - 0.2 MeV for
the $\Lambda$ separation energy
of $\hbox{}^{\ 6}_{\Lambda \Lambda} \hbox{He}$,
even when a rather compact $(3q)$ size of $b=0.6$ fm is assumed
as in our quark-model interactions.
Taking all of these effects into consideration, we can conclude
that the present results by fss2 are in good agreement
with the recent experimental value,
$\Delta B^{\rm exp}_{\Lambda \Lambda}
=1.01 \pm 0.20$ MeV, deduced from the Nagara event. \cite{TA01}

\bigskip

\begin{table}[t]
\caption{Comparison of $\Delta B_{\Lambda \Lambda}$ values
in MeV, predicted by various $\Lambda \Lambda$ interactions
and $V_{\Lambda N}$ potentials.
The $\Lambda \Lambda$ potential $V_{\Lambda \Lambda}$(Hiyama) is
the three-range Gaussian potential
used in Ref.~\protect\citen{HI97}.
FSS and fss2 use the $\Lambda \Lambda$ RGM $T$-matrix
in the free space.
The heading $\varepsilon_{\Lambda \Lambda}$ stands 
for the $\Lambda \Lambda$ expectation value
determined self-consistently,
and $V_{\Lambda \Lambda}$(SB) the two-range Gaussian potential
given in the text. In $V_{\Lambda \Lambda}$(SB) only the $S$-wave
is used, while in the others converged results
with enough partial waves are given.
}
\renewcommand{\arraystretch}{1.2}
\setlength{\tabcolsep}{4mm}
\begin{center}
\begin{tabular}{cccccccc}
\hline
{}& \multicolumn{2}{c}{$V_{\Lambda \Lambda}$(Hiyama)}
& \multicolumn{2}{c}{FSS} & \multicolumn{2}{c}{fss2}
& $V_{\Lambda \Lambda}$(SB) \\
\hline
$V_{\Lambda N}$ & present & Ref.~\protect\citen{HI97}
 & $\Delta B_{\Lambda \Lambda}$ & $\varepsilon_{\Lambda \Lambda}$
 & $\Delta B_{\Lambda \Lambda}$ & $\varepsilon_{\Lambda \Lambda}$
 & ($S$-wave) \\
\hline
SB & 3.618 & $-$  &  3.650 & 5.131 & 1.411 & 5.942 & 1.902 \\
NS & 3.548 & 3.59 &  3.623 & 5.159 & 1.364 & 5.951 & 1.917 \\
ND & 3.181 & 3.10 &  3.231 & 4.482 & 1.288 & 5.231 & 1.636 \\
NF & 3.208 & 3.22 &  3.301 & 4.626 & 1.270 & 5.410 & 1.719 \\
JA & 3.370 & 3.44 &  3.468 & 4.908 & 1.306 & 5.705 & 1.833 \\
JB & 3.486 & 3.56 &  3.592 & 5.150 & 1.325 & 5.956 & 1.924 \\
\hline
\end{tabular}
\label{table9}
\end{center}
\end{table}

\section{Summary}

From the advent of Yukawa's meson theory, a huge amount of efforts
have been devoted to understand the fundamental
nucleon-nucleon ($NN$) interaction and related hadronic interactions.
The present view of these interactions are non-perturbative
realization of inter-cluster interactions, governed by the
fundamental theory of the strong interaction,
quantum chromodynamics (QCD), in which the gluons are 
the field quanta exchanged between quarks.
On this basis, meson ``theory'' can be understood as 
an effective description of quark-gluon dynamics
in the low-energy regime. The short-range part of
the $NN$ and hyperon-nucleon ($YN$) interactions are
still veiled with unaccessible mechanism of quark
confinement and multi-gluon effects.

Here we have applied a naive constituent quark model
to the study of baryon-baryon interactions, in which
some of the essential features of QCD characteristics
are explicitly incorporated in the non-relativistic framework.
For example, the color degree of freedom of quarks is explicitly
incorporated, and the full antisymmetrization of quarks is carried out
in the resonating-group (RGM) formalism.
The gluon exchange effect is represented in the form of the quark-quark
interaction, for which a color analogue of the Fermi-Breit (FB)
interaction is used with an adjustable parameter of the
quark-gluon coupling constant $\alpha_S$.
The confinement potential is a phenomenological $r^2$-type potential,
which does not contribute to the baryon-baryon interactions
in the present framework.
Since the meson-exchange effects are the non-perturbative
aspect of QCD, these are described by the effective meson
exchange potentials (EMEP) acting between quarks.
After several improvements, the most recent
model called fss2 has achieved very accurate description
of the $NN$ and $YN$ interactions. \cite{FU96a}$^{\hbox{-}}$\cite{FU02a}

We have extended this $(3q)$-$(3q)$ RGM study of the
the $NN$ and $YN$ interactions 
to the strangeness $S=-2,~-3$ and $-4$ sectors
without introducing any extra parameters, and have clarified
some characteristic features
of the $B_8 B_8$ interactions. \cite{FU01b}
The results seem to be reasonable,
if we consider, i) the spin-flavor $SU_6$ symmetry,
ii) the weak pion effect in the strangeness sector,
and iii) the effect of the flavor symmetry breaking.
These $B_8 B_8$ interactions are now used for the detailed study
of the few-body systems, as well as baryonic matter problems,
in various ways.
Here we discussed applications
of the $NN$, $YN$ and $YY$ interactions
to the Faddeev calculations of the three-nucleon bound state,
the $\Lambda NN$-$\Sigma NN$ system for the hypertriton,
the $\alpha \alpha \Lambda$ system
for $\hbox{}^9_{\Lambda}\hbox{Be}$,
and the $\Lambda \Lambda \alpha$ system
for $\hbox{}^{\ 6}_{\Lambda \Lambda} \hbox{He}$.
We find that fss2 gives the triton
binding energy large enough to compare with the experiment,
without reducing the deuteron $D$-state probability. \cite{TR02,PANIC02}
The charge root-mean-square radii
of $\hbox{}^3\hbox{H}$ and $\hbox{}^3\hbox{He}$ are
also correctly reproduced.
The application to the hypertriton calculation shows that
fss2 gives a reasonable result similar to the Nijmegen
soft-core model NSC89, \cite{NSC89} except for an appreciable
contributions of higher partial waves. \cite{hypt}
In the application to the $\alpha \alpha \Lambda$ system,
the $\alpha \alpha$ RGM kernel
with the three-range Minnesota force \cite{TH77} and
some appropriate $\Lambda N$ force generated
from the low-energy phase-shift behavior of fss2 yield
the the ground-state and excitation energies
of $\hbox{}^9_{\Lambda}\hbox{Be}$ within  100 - 200 keV accuracy. \cite{2al}
The weak $\Lambda \Lambda$ force $\Delta B^{\rm exp}_{\Lambda
\Lambda}=1.01 \pm 0.20$ MeV deduced from
the Nagara event \cite{TA01}
for $\hbox{}^{\ 6}_{\Lambda \Lambda} \hbox{He}$ is reasonably
reproduced by the Faddeev calculation
of the $\Lambda \Lambda \alpha$ system,
using the present $\Lambda \alpha$ $T$-matrix
and the full coupled-channel $\Lambda \Lambda$-$\Xi
N$-$\Sigma \Sigma$ $\widetilde{T}$-matrix of fss2.

It is important to note that
the newly developed three-cluster Faddeev formalism
using two-cluster RGM kernels
opens a way to solve few-baryon systems interacting by the
quark-model baryon-baryon interactions without spoiling the
essential features of the RGM kernels; i.e., the non-locality,
the energy dependence and the existence of the pairwise
Pauli-forbidden state.
It can also be used for the three-cluster systems
involving $\alpha$-clusters,
like the $\hbox{}^9_{\Lambda}\hbox{Be}$
and $\hbox{}^{\ 6}_{\Lambda \Lambda} \hbox{He}$ systems.
A nice point of this formalism is that the
underlying $NN$ and $YN$ interactions are more directly
related to the structure of the hypernuclei
than the models assuming simple two-cluster
potentials. In particular, we have found that
the model fss2 yields a realistic description of many
three-body systems including the three-nucleon bound state,
the hypertriton, $\hbox{}^9_{\Lambda}\hbox{Be}$
and $\hbox{}^{\ 6}_{\Lambda \Lambda} \hbox{He}$. 

\section*{Acknowledgements}

This work was supported by Grants-in-Aid for Scientific
Research (C) (Nos. 15540270, 15540284 and 15540292) and
for Young Scientists (B) (No.~15740161) from
the Japan Society for the Promotion of Science (JSPS).

%

\end{document}